\documentclass[final,3p,times]{elsarticle}

\usepackage[]{todo}

\usepackage{latexsym}
\usepackage{amssymb}
\usepackage{amsthm}
\usepackage[cmex10]{amsmath}
\usepackage{graphicx}
\usepackage{float}
\usepackage{subfig}
\usepackage[]{placeins}
\usepackage[]{algorithm}
\usepackage{algpseudocode}
\usepackage{multirow}
\usepackage[subnum]{cases}
\usepackage{tabularx}

\usepackage[]{amsrefs} % without, The elsarticle class will remove the bibliography heading by declaring \let\bibsection\relax
\renewcommand{\algorithmicforall}{\textbf{for each}}

\makeatletter
\renewcommand\appendix{\par
  \setcounter{section}{0}%
  \setcounter{subsection}{0}%
  \setcounter{equation}{0}%
  \setcounter{table}{0}%------------ << add
  \setcounter{figure}{0}%----------- << add
  \gdef\theequation{\@Alph\c@section.\arabic{equation}}%
  \gdef\thefigure{\@Alph\c@section.\arabic{figure}}%
  \gdef\thetable{\@Alph\c@section.\arabic{table}}%
  \gdef\thesection{\Alph{section}}%
  \@addtoreset{equation}{section}%
  \@addtoreset{table}{section}%----- << add
  \@addtoreset{figure}{section}%---- << add
}
\makeatother

\begin{document}
\begin{frontmatter}
\title{Parallel mining of time--faded heavy hitters}
\author [unile] {Massimo~Cafaro\corref{cor1}}
\ead{massimo.cafaro@unisalento.it}
\cortext[cor1]{Corresponding author}
\author [unile] {Marco Pulimeno}
\ead{marco.pulimeno@unisalento.it}
\author [unile] {Italo Epicoco}
\ead{italo.epicoco@unisalento.it}
\address[unile]{University of Salento, Lecce, Italy}

\begin{abstract} We present PFDCMSS, a novel message--passing based parallel algorithm for mining time--faded heavy hitters. The algorithm is a parallel version of the recently published FDCMSS sequential algorithm. We formally prove its correctness by showing that the underlying data structure, a sketch augmented with a Space Saving stream summary holding exactly two counters, is mergeable. Whilst mergeability of traditional sketches derives immediately from theory, we show that merging our augmented sketch is non trivial. Nonetheless, the resulting parallel algorithm is fast and simple to implement. To the best of our knowledge, PFDCMSS is the first parallel algorithm solving the problem of mining time--faded heavy hitters on message--passing parallel architectures. Extensive experimental results confirm that PFDCMSS retains the extreme accuracy and error bound provided by FDCMSS whilst providing excellent parallel scalability.
\end{abstract}


\theoremstyle{plain}% default
\newtheorem{thm}{Theorem}
\newtheorem{lem}[thm]{Lemma}
\newtheorem{prob}{Problem}
\newdefinition{rmk}{Remark}
\newproof{pf}{Proof}
 \newproof{pot}{Proof of Theorem \ref{thm2}}
\newtheorem{prop}[thm]{Proposition}
\newtheorem*{cor}{Corollary}
\newtheorem{corollary}[thm]{Corollary}
\newtheorem{observation}[thm]{Observation}
\newdefinition{defn}{Definition}
\newtheorem{conj}{Conjecture}
\newtheorem{exmp}{Example}
\theoremstyle{remark}
\newtheorem*{rem}{Remark}
\newtheorem*{note}{Note}
\newtheorem{case}{Case}
\newtheorem{claim}[thm]{Claim}
\newtheorem{fact}[thm]{Fact}
\newtheorem{assumption}[thm]{Assumption}

%\tableofcontents
\end{frontmatter}

%\tableofcontents

\section{Introduction}
\label{intro}

In this paper we deal with the problem of mining in parallel time--faded heavy hitters (also called frequent items), and we present  PFDCMSS, a novel message--passing based parallel algorithm which is a parallel version of the recently published FDCMSS sequential algorithm \cite{Cafaro-Pulimeno-Epicoco-Aloisio}.
 
Mining of heavy hitters in a data stream has been thoroughly studied, and the problem is regarded as one of the most important in the streaming algorithms literature. Depending on the particular application, the problem is reported in the literature as \textit{hot list analysis} \cite{Gibbons}, market basket analysis \cite{Brin} and \textit{iceberg query} \cite{Fang98computingiceberg}, \cite{Beyer99bottom-upcomputation}. 

Even though there are many possible applications, we recall here some of the most important contexts to which the problem has been successfully applied: network traffic analysis \cite{DemaineLM02},  \cite{Estan}, \cite{Pan}, analysis of web logs \cite{Charikar}, Computational and theoretical Linguistics \cite{CICLing}.

All of the algorithms for detecting heavy hitters can be classified as being either \emph{counter} or \emph{sketch} based, the difference being that counter--based algorithms rely on a set of counters which are used to keep track of stream items, whilst sketch--based algorithms monitor the data stream by using a sketch data structure, often a bi-dimensional array data structure containing a counter in each cell. Stream items are mapped by hash functions to corresponding cells in the sketch. The former algorithms (counter--based) are deterministic, whilst the latter (sketch--based) are probabilistic.

Regarding counter--based algorithms, the first sequential algorithm has been designed by Misra and Gries \cite{Misra82}. Their algorithm was rediscovered, independently, about twenty years later by Demaine et al. \cite{DemaineLM02} (this algorithm is known in the literature as the \emph{Frequent} algorithm) and Karp et al. \cite{Karp}. Among the developed counters--based algorithms we recall here  \emph{Sticky Sampling}  and \emph{Lossy Counting} \cite{Manku02approximatefrequency}, and \emph{Space Saving} \cite{Metwally2006}. Sketch--based solutions include \emph{CountSketch} \cite{Charikar}, \emph{Group Test} \cite{Cormode-grouptest}, \emph{Count-Min} \cite{Cormode05} and \emph{hCount} \cite{Jin03}.

Relevant parallel algorithms include \cite{cafaro-tempesta}, \cite{Cafaro-Pulimeno} and \cite{Cafaro-Pulimeno-Tempesta} which are  message-passing based parallel versions of the Frequent and Space Saving algorithms. Shared-memory algorithms have been designed as well, including a parallel version of Frequent \cite{Zhang2013}, a parallel version of Lossy Counting \cite{Zhang2012}, and parallel versions of Space Saving \cite{Roy2012} \cite{Das2009}. Recent shared-memory parallel algorithms for heavy hitters were recently proposed in \cite{Tangwongsan2014}. Finally, accelerator based algorithms exploiting a GPU (Graphics Processing Unit) include \cite{Govindaraju2005} and \cite{Erra2012}. Regarding related work, i.e., parallel algorithms specifically designed to solve the problem of mining time--faded heavy hitters, we are not aware of any other algorithm: to the best of our knowledge, ours is the first parallel algorithm solving the problem on message--passing parallel architectures.

In this paper, we are concerned with the problem of detecting in parallel heavy hitters in a stream with the additional constraint that recent items must be weighted more than former items. The underlying assumption is that, in some applications, recent data is certainly more useful and valuable than older, stale data. Therefore, each item in the stream has an associated timestamp that will be used to determine its weight. In practice, instead of estimating items' frequencies, we are required to estimate items' \emph{decayed frequencies}. 

This paper is organized as follows. We recall in Section \ref{definitions} preliminary definitions and concepts that will be used in the rest of the manuscript. We present in Section \ref{alg} our PFDCMSS algorithm and formally prove in Section \ref{correctness} its correctness. Next, we provide extensive experimental results in Section \ref{results}, showing that PFDCMSS retains the extreme accuracy and error bound provided by the sequential FDCMSS whilst providing excellent parallel scalability. Finally, we draw our conclusions in Section \ref{conclusions}.

\section{Preliminary definitions}
\label{definitions}

In this Section we introduce preliminary definitions and the notation used throughout the paper. We deal with an input data stream $\sigma$ consisting of a sequence of $n$ items drawn from a universe $\mathcal{U}$; without loss of generality, let $m$ be the number of distinct items in $\sigma$ i.e., let $\mathcal{U}=\{1,2,\ldots,m\}$, which we will also denote as $[m]$. Let $f_i$ be the frequency of the item $i \in \mathcal{U}$ (i.e., its number of occurrences in $\sigma$), and denote the frequency vector by $\textbf{f} = (f_1,\ldots,f_m)$. Moreover, let $0 < \phi < 1$ be a support threshold, $0 < \epsilon < 1$ a tolerance such that $\epsilon < \phi$ and denote the 1-norm of $\textbf{f}$ (which represents the total number of occurrences of all of the stream items) by $||\textbf{f}||_1$. 

In this paper, we are concerned with the problem of detecting in parallel frequent items in a stream with the additional constraint that recent items must be weighted more than former items. The underlying assumption is that, in some applications, recent data is certainly more useful and valuable than older, stale data. Therefore, each item in the stream has an associated timestamp that will be used to determine its weight. In practice, instead of estimating frequencies, we are required to estimate \emph{decayed frequencies}. Two different models have been proposed in the literature: the \emph{sliding window} and the \emph{time fading} model. PFDCMSS works in the latter model. Furthermore, even though the basic ideas underlining the algorithm are also appropriate for an online distributed setting, here we are assuming that the entire dataset is available for offline processing.

The time fading model \cite{recent-freq-items} \cite{exp-decay} \cite{Chen-Mei} does not use a window sliding over time; freshness of more recent items is instead emphasized by \emph{fading} the frequency count of older items. This is achieved by computing the item's \textit{decayed frequency} through the use of a decay function that assign greater weight to more recent occurrences of an item than to older ones: the older an occurrences is, the lower its decayed weight.

\begin{defn}
\label{decay-function}
Let $w(t_i,t)$ be a decayed function which computes the decayed weight at time $t$ for the occurrence of item $i$ arrived at time $t_i$. A decayed function must satisfy the following properties: 
\begin {enumerate}
\item $w(t_i,t) = 1$ when $t_i = t$ and $0 \leq w(t_i,t) \leq 1$ for all $t > t_i$;
\item $w$ is a monotone non-increasing function as time $t$ increases, i.e., $t' \geq t \implies w(t_i, t') \leq w(t_i, t)$.
\end{enumerate}
\end{defn}

Related work has mostly exploited \textit{backward decay} functions, in which the weight of an item is a function of its age, $a$, where the age at time $t > t_i$ is simply $a = t-t_i$. In this case, $w(t_i, t)$ is given by $w(t_i, t) = \frac{h(t-t_i)}{h(t-t)}=\frac{h(t-t_i)}{h(0)}$, where $h$ is a positive monotone non-increasing function.\\
The term backward decay stems from the aim of measuring from the current time back to the item's timestamp. Prior algorithms and applications have been using backward exponential decay functions such as $h(a) = e^{-\lambda a}$, with $\lambda > 0$ as decaying factor. 

In our algorithm, we use instead a forward decay function, defined as follows (see \cite{forward-decay} for a detailed description of the forward decay approach). Under forward decay, the weight of an item is computed on the amount of time between the arrival of an item and a fixed point $L$, called the \textit{landmark} time, which, by convention, is some time earlier than the timestamps of all of the items. The idea is to look forward in time from the landmark to see an item, instead of looking backward from the current time.

\begin{defn}
	Given a positive monotone non-decreasing function $g$, and a landmark time $L$, the forward decayed weight of an item $i$ with arrival time $t_i > L$ measured at time $t \geq t_i$ is given by $w(t_i, t) = \frac{g(t_i-L)}{g(t-L)}$.
\end{defn}

The denominator is used to normalize the decayed weight so that $w(t_i, t)$ is always less than or equal to 1 as requested by Definition~\ref{decay-function}.

\begin{defn}
\label{item-decayed-count}
The \textit{decayed frequency} of an item $v$ in the input stream $\sigma$, computed at time $t$, is given by the sum of the decayed weights of all the occurrences of $v$ in $\sigma$: $f_v(t) = \sum_{v_i = v} w(t_i,t)$.
\end{defn}

\begin{defn}
\label{decayed-count}
The \textit{decayed count} at time $t$, $C(t)$, of a stream $\sigma$ of $n$ items is the sum of the decayed weights of all the items occurring in the stream: $C(t)=\sum_{i=1}^n w(t_i, t)$.
\end{defn}

The Approximate Time--Faded Heavy Hitters (ATFHH) problem is formally stated as follows.

\begin{prob} Approximate Time--Faded Heavy Hitters. Given a stream $\sigma$ of items with an associated timestamp, a threshold $0 < \phi < 1$ and a tolerance $0 < \epsilon < 1$ such that $\epsilon < \phi$, and letting $g$ be a decaying function used to determine the decayed frequencies and $t$ be the query time, return the set of items $F$, so that:

\begin{itemize}
\item $F$ contains all of the items $v$ with decayed frequency at time $t$ $f_v(t) > \phi C(t)$ (decayed frequent items);
 
\item $F$ does not contain any item $v$ such that $f_v(t) \leq (\phi-\epsilon) C(t)$. 
\end{itemize}

\end{prob}

In the following, when clear from the context, the query time shall be considered an implicit parameter, so we write $f_v$ and $C$ instead of $f_v(t)$ and $C(t)$. The algorithm presented makes use of a Count--Min sketch data structure augmented by a Space Saving summary associated to each sketch cell. In the following, we recall the main properties of the Count--Min and the Space Saving algorithms in the case of non decaying frequencies, but the same properties also hold in a time-fading context.

Count--Min is based on a sketch whose dimensions are derived by the input parameters $\epsilon$, the error, and $\delta$, the probability of failure. In particular, for Count--Min $d=\lceil \ln 1/\delta \rceil$ is the number of rows in the sketch and $w=\lceil e/\epsilon \rceil$ is the number of columns. Every cell in the sketch is a counter, which is updated by hash functions. By using this data structure, the algorithm solves with probability greater than or equal to 1 - $\delta$ the \textit{frequency estimation} problem for arbitrary items. The algorithm may also be extended to solve the \textit{approximate frequent items} problem as well, by using an additional heap data structure which is updated each time a cell is updated. Since in Count-Min the frequencies stored in the cells overestimate the true frequencies, a point query for an arbitrary item simply inspects all of the $d$ cells in which the item is mapped to by the corresponding hash functions and returns the minimum of those $d$ counters. 

Space Saving is a counter-based algorithms solving the heavy hitters problem. It makes use of a stream summary data structure composed by a given number of counters $k \ll n$, $n$ being the length of the stream. Each counter monitors an item in the stream and tracks its frequency. A substitution strategy is used when the algorithm processes an item not already monitored and all of the counters are occupied.

Let $\sigma$ be the input stream and denote by $\mathcal{S}$ the summary data structure of $k$ counters used by the Space Saving algorithm. Moreover, denote by $\left|\mathcal{S}\right|$ the sum of the counters in $\mathcal{S}$, by $f_v$ the exact frequency of an item $v$ and by $\hat{f}_v$ its estimated frequency, let $\hat{f}^{min}$ be the minimum frequency in $\mathcal{S}$. If there exist at least one counter not monitoring any item, $\hat{f}^{min}$ is zero.

Finally, denote by $\textbf{f} = (f_1,\ldots,f_m)$ the frequency vector. The following relations hold (as shown in \cite{Metwally2006}):

\begin{equation}
\label{ss1}
\left|\mathcal{S}\right| = ||\textbf{f}||_1,
\end{equation}

\begin{equation}
\label{ss2}
\hat{f}_v - \hat{f}^{min} \leq f_v \leq \hat{f}_v,  \qquad v \in \mathcal{S},
\end{equation}

\begin{equation}
\label{ss3}
f_v  \leq \hat{f}^{min}, \qquad \hspace{13mm} v \notin \mathcal{S},
\end{equation}

\begin{equation}
\label{ss4}
\hat{f}^{min}  \leq \left\lfloor\frac{||\textbf{f}||_1}{k}\right\rfloor.
\end{equation}

Therefore, it holds that

\begin{equation}
\label{ss5}
\hat{f}_v - f_v \leq \hat{f}^{min} \leq \left\lfloor\frac{||\textbf{f}||_1}{k}\right\rfloor, \hspace{3mm} v \in \mathcal{U}.
\end{equation}

\section{The algorithm}
\label{alg}

In this section, we start by recalling our sequential algorithm FDCMSS \cite{Cafaro-Pulimeno-Epicoco-Aloisio}. The key data structure is an augmented Count--Min sketch $D$, whose dimensions $d$ (rows) and $w$ (columns) are derived by the input parameters $\epsilon$, the error, and $\delta$, the probability of failure. Whilst every cell in an ordinary CM sketch contains a counter used for frequency estimation, in our case a cell holds a Space Saving stream summary with exactly two counters. The idea behind the augmented sketch is to monitor the time--faded items that the sketch hash functions map to the corresponding cells by an instance of Space Saving with two counters, so that for a given cell we are able to determine a \textit{majority item candidate} with regard to the sub-stream of items falling in that cell. 

Indeed, by using a data structure $\mathcal{S}$ with two counters in each cell, and letting $C_{i,j}$ denote the total decayed count of the items falling in the cell $D[i][j]$, the majority item is, if it exists, the item whose decayed frequency is greater than $\frac{C_{i,j}}{2}$. The corresponding majority item candidate in the cell is the item monitored by the Space Saving counter whose estimated decayed frequency is maximum. We have proved that, with high probability, if a time-faded item is frequent, then, in at least one of the sketch cells where it is mapped, it is a majority item with regard to the sub-stream of items  falling in the same cell. Therefore, our algorithm will detect it. 

\begin{thm}
If an item $i$ is frequent, then it appears as a majority item candidate in at least one of the $d$ cells in which it falls, with probability greater than or equal to $1 - (\frac{1}{2 \phi w})^d$. 
\end{thm}

Regarding the error bound of our algorithm, let $f_i$ be the exact decayed frequency of item $i$ in the stream $\sigma$ and $\hat{f}_i$ be the estimated decayed frequency of item $i$ returned by FDCMSS. Let $C$ be the total decayed count of all of the items in the stream. We have proved the following error bound.

\begin{thm}
\label{error-bound}
$\forall u \in [m]$, $\hat{f}_u$ estimates the exact decayed count $f_u$ of $u$ at query time  with error less than $\epsilon C$ and probability greater than $1-\delta$.
\end{thm}

The proofs of aforementioned theorems can be found in  \cite{Cafaro-Pulimeno-Epicoco-Aloisio}. 

The algorithm's initialization requires as input parameters $\epsilon$, the error; $\delta$, the probability of failure; and $\phi$, the support threshold. The initialization returns a sketch $D$. The procedure starts deriving $d=\lceil \ln 1/\delta \rceil$, the number of rows in the sketch and $w=\lceil \frac{e}{2\epsilon} \rceil$, the number of columns in the sketch. Then, for each of the $d*w$ cells available in the sketch $D$ we allocate a data structure $\mathcal{S}$ with two Space Saving counters $c_1$ and $c_2$. Given a counter $c_j, j=1,2$, we denote by $c_j.i$ and $c_j.f$ respectively the counter's item and its estimated decayed frequency. Finally, we set the support threshold to $\phi$, select $d$ pairwise independent hash functions $h_1,\ldots,h_d:[m] \rightarrow [w]$, mapping $m$ distinct items into $w$ cells, and initialize the \textit{count} variable, representing the total decayed count of all of the items in the stream, to zero.

Updating the sketch upon arrival of a stream item $i$ with timestamp $t_i$, shown in pseudo-code as Algorithm \ref{process}, requires computing $x$, which is the non normalized forward decayed weight of the item, and incrementing \textit{count} by $x$. Then, we update the $d$ cells in which the item is mapped to by the corresponding hash functions $h_j(x), j=1,\dots,d$ by using the Space Saving item update procedure. 

\begin{algorithm}
\begin{algorithmic}[1]
\Require $i$, an item; $t_i$, timestamp of item $i$; D, sketch data structure
\Ensure update of sketch related to item $i$; update the local total decayed count.
\Procedure {process}{$i, t_i, D$}
\State $x \leftarrow g(t-t_i)$
\Comment{compute the non normalized decayed weight of item $i$}
\State $lCount \leftarrow lCount + x$
\Comment {update local total decayed count}
\For{$j=1$ to $d$}
	\State $\mathcal{S} \leftarrow D[j][h_j(i)]$
	\State \Call{SpaceSavingUpdate}{$\mathcal{S}, i, x$}
	\Comment {update the sketch}
\EndFor
\EndProcedure
\caption{Process}
\label{process}
\end{algorithmic}
\end{algorithm}

Let $\mathcal{S}$ denote the Space Saving stream summary data structure with two counters corresponding to the cell to be updated. Updating $\mathcal{S}$ upon arrival of an item works as follows. When processing an item which is already monitored by a counter, its estimated frequency is incremented by the non normalized weight $x$. When processing an item which is not already monitored by one of the available counters, there are two possibilities. If a counter is available, it will be in charge of monitoring the item, and its estimated frequency is set to the non normalized weight $x$.  Otherwise, if all of the counters are already occupied (their frequencies are different from zero), the counter storing the item with minimum frequency is incremented by the non normalized weight $x$. Then, the monitored item is evicted from the counter and replaced by the new item. This happens since an item which is not monitored can not have a frequency greater than the minimal frequency.

PFDCMSS, the parallel version of our sequential algorithm, works as follows. We assume the offline setting in which the stream items have been stored as a static dataset along with the corresponding timestamps. It is worth noting here immediately that our algorithm works in the streaming (online) setting as well. Indeed, in the former case (offline setting) we partition the input dataset and timestamps using a simple 1D block-based domain decomposition among the available $p$ processes and then process in parallel the sub-streams assigned to the processes using Algorithm \ref{process}. In the latter case (online setting), we have instead $p$ distributed sites, each handling a different stream $\sigma_i, i = 1,\ldots,p$ processed again using Algorithm \ref{process}. 

In the parallel version, once the sub-streams have been processed, one of the processes is in charge of determining the time--faded heavy hitters. In order to do so, all of the processes engage in a parallel reduction in which their sketches are merged into a global sketch which preserves all of the information stored in the local sketches. This sketch is then queried and the time--faded heavy hitters are returned.  

In the distributed setting, one of the sites may act as a centralized coordinator or there can be another different site taking this responsibility. The coordinator broadcasts, when required, a "query" message to the $p$ sites, which then temporarily stop processing their sub-streams, and engage in the sketch merge procedure. We can imagine the distributed sites as being multi-threaded processes, in which one thread executes Algorithm \ref{process}, temporarily stops when a query message is received from the coordinator, creates a copy of its local sketch and then resume stream processing whilst another thread engages in the distributed sketch merging procedure using the sketch copy.

In order to retrieve the time--faded heavy hitters, a query can be posed when needed. The query, shown in pseudo-code as Algorithm \ref{query}, starts by determining the global decayed count for the whole stream $\sigma$. This requires a parallel reduction in which the local decayed counts are summed. It is worth noting here that the global decayed count is still non normalized; the normalization occurs dividing by $g(t-L)$, where $t$ is the query time and $L$ denotes the landmark time. Then, we build, through a user's defined parallel reduction, a global sketch $G$ which is obtained by merging the local sketches. To do so, each process invokes a parallel reduction by using the MergeSketch operator shown in pseudo-code as Algorithm~\ref{ParallelSketchReductionOperator}. 

\begin{algorithm}[t]
\begin{algorithmic}[1]
\Require $t$, query time; $D$, process' sketch
\Ensure set of frequent items
\Procedure {query}{$t$}
\State $gCount \leftarrow$ \Call{ParallelReduction}{$lCount, \textsc{Sum}$}
\State $gCount \leftarrow \frac{gCount}{g(t-L)}$
\State $G \leftarrow$ \Call{ParallelReduction}{$D, \textsc{MergeSketch}$}
\State $F=\emptyset$
\ForAll{$\mathcal{S}_{ij} \in G$}
		\State let $c_1$ and $c_2$ be the counters in $\mathcal{S}_{ij}$ 
		\State $c_m \leftarrow \Call{argmax}{c_1, c_2}$
		\Comment {$c_m$ the counter with maximum decayed count}
		\If{$\frac{c_m.f}{g(t-L)}  > \phi *gCount$}	
			\State $p \leftarrow \Call{PointEstimate}{c_m.i, t}$
			\If{$p > \phi *gCount$}
				\State $F \leftarrow F \cup \{(c_m.i, p)\}$
			\EndIf
		\EndIf
\EndFor
\State \Return $F$
\EndProcedure
\caption{Query}
\label{query}
\end{algorithmic}
\end{algorithm}

\begin{algorithm}
\begin{algorithmic}[1]
\Require $D_1, D_2$: sketches to be merged.
\Ensure $G$, the merged sketch
\Procedure {MergeSketch}{$D_1, D_2$}
\ForAll{$\mathcal{S}^1_{ij} \in D_1, \mathcal{S}^2_{ij} \in D_2$}
		\State $m_1 \leftarrow $ \Call {min}{$c^1_1.f, c^1_2.f$}
		\Comment{$m_1$, the minimum of counters' frequency in $\mathcal{S}^1_{ij}$}
		\State $m_2 \leftarrow $ \Call {min}{$c^2_1.f, c^2_2.f$}
		\Comment{$m_2$, the minimum of counters' frequency in $\mathcal{S}^2_{ij}$}
\ForAll{$c_{s_1} \in \mathcal{S}_1$}
	\State $c_{s_2} \leftarrow \Call {Find}{\mathcal{S}_2, c_{s_1}.i}$
	\If{$c_{s_2}$}
		\State $c_{s_c}.f \leftarrow c_{s_1}.f + c_{s_2}.f$
		\State \Call {Delete}{$\mathcal{S}_2, c_{s_2}$}
	\Else
		\State $c_{s_c}.f \leftarrow c_{s_1}.f + m_2$
	\EndIf
	\State $c_{s_c}.i \leftarrow c_{s_1}.i$
	\State \Call {Insert}{$\mathcal{S}_C, c_{s_c}$}
\EndFor
\ForAll{$c_{s_2} \in \mathcal{S}_2$}
	\State $c_{s_c}.i \leftarrow c_{s_2}.i$
	\State $c_{s_c}.f \leftarrow c_{s_2}.f + m_1$
	\State \Call {Insert}{$\mathcal{S}_C, c_{s_c}$}
\EndFor
		\State \Call {Purge}{$\mathcal{S}_C$}
		\Comment {$\mathcal{S}_C$ now contains $2$ counters with the greatest frequencies}
		\State $G[i][j] \leftarrow \mathcal{S}_C$
\EndFor
\State \Return $G$
\EndProcedure
\caption{MergeSketch}
\label{ParallelSketchReductionOperator}
\end{algorithmic}
\end{algorithm}

The sketches are reduced as follows: for every corresponding cell in two  sketches to be merged, the hosted Space Saving summaries are merged following the steps described in \cite{Cafaro-Pulimeno-Tempesta}, i.e., building a temporary summary $\mathcal{S}_C$ consisting of all of the items monitored by both $\mathcal{S}_1$ and $\mathcal{S}_2$. To each item in $\mathcal{S}_C$ is assigned a decayed frequency computed as follows: if an item is present in both $\mathcal{S}_1$ and $\mathcal{S}_2$, its frequency is the sum of the its corresponding frequencies in each summary; if the item is present only in one of either $\mathcal{S}_1$ or $\mathcal{S}_2$, its frequency is incremented by the minimum frequency of the other summary. At last, in order to derive the merged summary, we take only the $2$ items in $\mathcal{S}_C$ with the greatest frequencies and discard the others. 

It is worth noting here that the sum of the counters in the stream summary data structure $\mathcal{S}$ related to a given cell $D[i][j]$ is equal to the value that the Count--Min sketch--based algorithm would store in the counter variable corresponding to that cell, i.e., the 1-norm of the frequency vector corresponding to the sub--stream falling in the cell through the pairwise independent hash functions. Thus, an augmented sketch is equivalent, from this perspective, to a Count--Min sketch and this property is preserved by the merge procedure. From now on we will call this property 1-\textit{norm equivalence}.

However, merging Count--Min sketches simply requires adding the corresponding cells' counters. Indeed, via linearity, the sum of sketches is equal to the sketch of the sums. Instead, in our case, we need an ad hoc procedure in order to correctly merge the two Space Saving stream summaries hosted by the corresponding cells so that 1-norm equivalence property is preserved. Nonetheless, the augmented sketch which results from our parallel merge reduction is 1-norm equivalent to the Count--Min sketch obtained by summing the Count--Min sketches corresponding to our augmented sketches which are the input of the parallel merge reduction.

Once the global sketch is obtained, the query procedure initializes $F$, an empty set, and then it inspects each of the $d*w$ cells in the sketch $D$. For a given cell, we determine $c_m$, the counter in the data structure $\mathcal{S}$ with maximum decayed count. We normalize the decayed count stored in $c_m$ dividing by $g(t-L)$, and then compare this quantity with the threshold given by $\phi*gCount$ ($gCount$ being the normalized global decayed count). If the normalized decayed frequency is greater, we pose a point query for the item $c_m.i$, shown in pseudo-code as Algorithm \ref{estimate}. If $p$, the returned value, is greater than the threshold $\phi*gCount$, then we insert in $F$ the pair $(c_m.i,p)$.

\begin{algorithm}
\begin{algorithmic}[1]
\Require $j$, an item; $t$, query time
\Ensure estimation of item $j$ decayed count;
\Procedure {pointestimate}{$j, t$}
\State $answer \leftarrow \infty$
\For{$i=1$ to $d$}
	\State $\mathcal{S} \leftarrow G[i][h_i(j)]$
	\Comment{let $c_1$ and $c_2$ be the counters in $\mathcal{S}$}
	\If{$j == c_1.i$}
		\State $answer \leftarrow \Call{min}{answer, c_1.f}$
	\ElsIf{$j == c_2.i$}
			\State $answer \leftarrow \Call{min}{answer, c_2.f}$
		\Else
			\State $m \leftarrow \Call{min}{c_1.f, c_2.f}$
			\State $answer \leftarrow \Call{min}{answer, m}$
		\EndIf
\EndFor
\State \Return $\frac{answer}{g(t-L)}$
\EndProcedure
\caption{PointEstimate}
\label{estimate}
\end{algorithmic}
\end{algorithm}

The point query for an item $j$ returns its estimated decayed frequency. After initializing the \textit{answer} variable to infinity, we inspect each of the $d$ cells in which the item is mapped to by the corresponding hash functions, to determine the minimum decayed frequency of the item. In each cell, if the item is stored by one of the Space Saving counters, we set \textit{answer} to the minimum between \textit{answer} and the corresponding counter's decayed frequency. Otherwise (none of the two counters monitors the item $j$), we set \textit{answer} to the minimum between \textit{answer} and the minimum decayed frequency stored in the counters. Since the frequencies stored in all of the counters of the sketch are not normalized, we return the normalized frequency \textit{answer} dividing by $g(t-L)$.

At the end of the query procedure the set $F$ is returned.

\section{Correctness}
\label{correctness}

Here, we prove that our algorithm correctly merges two FDCMSS sketches. The merge procedure preserves all of the properties of the sketch, including the fact that, considering the sum of the Space Saving counters in each sketch cell, an FDCMSS sketch is 1-norm equivalent to the classical Count--Min sketch. 

It is worth noting here that we would obtain a correct result by using the merge procedure presented in \cite{Cafaro-Pulimeno-Tempesta} to combine the Space Saving summaries stored in the corresponding sketch cells, but we also want to impose 1-norm equivalence, i.e., the additional condition that the sum of counters' values in each merged cell always reflects the total decayed count of the items which fell in the corresponding cells.

Indeed, in \cite{Cafaro-Pulimeno-Tempesta} we showed how to merge Space Saving stream summaries in parallel. However, we have proved that our merge procedure satisfies the Space Saving properties described by eq. \ref{ss2}-\ref{ss5}, and the following relaxed version of eq. \ref{ss1}:

\begin{equation}
\label{ss6}
\left|\mathcal{S}\right| \leq ||\textbf{f}||_1,
\end{equation}

As shown in Theorem~\ref{th_merge}, which is the main result of this section, it turns out that $k = 2$ counters (i.e., majority item mining) is a special case: when the Space Saving summaries to be merged hold two counters, than the property in eq. \ref{ss1} holds for the merged summary in its original form, that is \text{$\left|\mathcal{S}\right| = ||\textbf{f}||_1$}, without modifying the merge procedure designed in \cite{Cafaro-Pulimeno-Tempesta}. 

\begin{thm}
\label{th_merge}
The parallel merge algorithm provides an augmented sketch that preserves all of the properties of a FDCMSS sketch.
\end{thm}

\begin{proof}
	
The correctness of the parallel FDCMSS sketch merge algorithm derives from the correctness of the Space Saving merge procedure, already shown in \cite{Cafaro-Pulimeno-Tempesta}. It remains to show that, when looking to the sum of the Space Saving counters associated to each cell, the merged augmented sketch is still 1-norm equivalent to a Count--Min sketch, that is, the sum of the counters values is equal to the decayed count of all the items fallen in that cell.

Let us recall the merge algorithm for Space Saving summaries introduced in \cite{Cafaro-Pulimeno-Tempesta}. We will use the multiset notation, thus let us rewrite the properties of a Space Saving summary stated in equations \ref{ss1}-\ref{ss4}, this time with reference to multisets. Indeed, we model the input stream as a multiset (also called a \emph{bag}), which essentially is a set where the duplication of elements is allowed. We shall use a calligraphic capital letter to denote a multiset, and the corresponding capital Greek letter to denote its \textit{underlying} set. In particular, we extend the traditional notion of multiset as follows. Instead of considering an indicator function which returns the multiplicity of an item, we use a function providing the decayed frequency of that item. Therefore, summing over all of the items we obtain the total decayed count in place of the cardinality of the multiset. 

\begin{defn}
\label{multiset}
A decayed multiset $\mathcal{N}=(N, f_{\mathcal{N}})$ is a pair where $N$ is some set, called the underlying set of elements, and $f_{\mathcal{N}}: N \rightarrow \mathbb{R}$ is a function which provides the \textit{decayed frequency} for each $x \in N$ according to Definition \ref{item-decayed-count}.
%The generalized indicator function of $\mathcal{N}$ is 
%\begin{equation}
%\label{eq01}
%I_\mathcal{N} (x) := \left\{ {\begin{array}{*{20}c}
%   {f_{\mathcal{N}}(x)} & {x \in N} , \\
%   0 & {x \notin N},  \\
% \end{array} }\right.
% \end{equation}

The decayed count of $\mathcal{N}$ is expressed by

\begin{equation}
\label{eq02}
\left\vert{\mathcal{N}}\right\vert :=  \sum\limits_{x \in N} {f_\mathcal{N} (x)},
\end{equation}

whilst the cardinality of the underlying set $N$ is

\begin{equation}
\label{eq03}
\left\vert{N}\right\vert := Card(N) = \sum\limits_{x \in N} {1}.
\end{equation}
\end{defn}

From now on, when referring to either the exact or estimated frequency of an item, we shall mean the item's exact or estimated decayed frequency. Recall that our Space Saving stream summary data structure uses exactly $k = 2$ counters, and let $\mathcal{N} = (N, f_{\mathcal{N}})$ be the input decayed multiset, $\mathcal{S} = (\Sigma, \hat{f_{\mathcal{S}}})$ the decayed multiset of all of the monitored items and their respective counters at the end of the sequential Space Saving algorithm's execution, i.e., the algorithm's summary data structure. Let  $\left\vert \mathcal{S} \right\vert$ be the sum of the frequencies stored in the counters, $f_\mathcal{N}(e)$ the exact frequency of an item $e$, $\hat{f_\mathcal{S}}(e)$ its estimated frequency and $\hat{f_\mathcal{S}}^{min}$ the minimum frequency in $\mathcal{S}$, where  $\hat{f_\mathcal{S}}^{min} = 0$ when  $\left\vert{\Sigma}\right\vert < 2$. Indeed, even though a summary data structure has exactly 2 counters, it may monitor less than 2 items, since an item is actually monitored if and only if its counter's frequency is different from zero. The following relations hold, for each item $e \in N$:

\begin{equation}
\label{eq04}
\left\vert{\mathcal{S}}\right\vert = \left\vert{\mathcal{N}}\right\vert,
\end{equation}

\begin{equation}
\label{eq05}
\hat{f_\mathcal{S}}(e) - \hat{f_\mathcal{S}}^{min} \leq f_\mathcal{N}(e) \leq \hat{f_\mathcal{S}}(e),  \qquad e \in \Sigma,
\end{equation}

\begin{equation}
\label{eq06}
f_\mathcal{N}(e)  \leq \hat{f_\mathcal{S}}^{min}, \qquad e \notin \Sigma,
\end{equation}

\begin{equation}
\label{eq07}
\hat{f_\mathcal{S}}^{min}  \leq \left\lfloor\frac{\left\vert{\mathcal{N}}\right\vert}{2}\right\rfloor.
\end{equation}
\\

Now, let $\mathcal{S}_1 = (\Sigma_1, \hat{f}_{\mathcal{S}_1})$ and $\mathcal{S}_2 = (\Sigma_2, \hat{f}_{\mathcal{S}_2})$ be two summaries related respectively to the input sub-arrays $\mathcal{N}_1 = (N_1, f_{\mathcal{N}_1})$ and $\mathcal{N}_2 = (N_2, f_{\mathcal{N}_2})$, with $\mathcal{N} = \mathcal{N}_1 \uplus \mathcal{N}_2 = (N, f_{\mathcal{N}})$. Let $\mathcal{S}_M = (\Sigma_M, \hat{f}_{\mathcal{S}_M})$ be the final merged summary. 

Theorem 3 in \cite{Cafaro-Pulimeno-Tempesta} states that if eqs. (\ref{eq05}) - (\ref{eq07}) hold for $\mathcal{S}_1$ and $\mathcal{S}_2$ and, if it is verified a relaxed version of eq. (\ref{eq04}), i.e., it holds that

\begin{equation}
\label{eq21}
\left\vert{\mathcal{S}_i}\right\vert \leq \left\vert{\mathcal{N}_i}\right\vert, \qquad i = 1,2
\end{equation}
\\
then these properties  continue to be true also for $\mathcal{S}_M$ (it is worth noting here that eq. (\ref{eq21}) also holds for summaries produced by the sequential Space Saving algorithm). The  authors show that this is enough to guarantee the correctness of the merge operation, but, in general \text{$\left\vert{\mathcal{S}_M}\right\vert \leq \left\vert{\mathcal{N}}\right\vert$}.

In order to obtain $\mathcal{S}_M$, we start combining $\mathcal{S}_1$ and $\mathcal{S}_2$ to obtain $\mathcal{S}_C$, and then, if $\left\vert\Sigma_C\right\vert > 2$, we take the two counters with the greatest frequency values in $\mathcal{S}_C$ in order to build $\mathcal{S}_M$, otherwise we return $\mathcal{S}_M = \mathcal{S}_C$.

We can express the combine operation as shown by the following equation:

\begin{equation}
\label{eq08}
\hat{f}_{\mathcal{S}_C}(e) = 
\left\lbrace 
\begin{array}{r} 
\vspace{0.3cm}
\hat{f}_{\mathcal{S}_1}(e) + \hat{f}_{\mathcal{S}_2}(e), \qquad e \in \Sigma_1 \cap \Sigma_2, \\
\vspace{0.3cm}
\hat{f}_{\mathcal{S}_1}(e) + \hat{f}_{\mathcal{S}_2}^{min}, \qquad e \in \Sigma_1 \setminus \Sigma_2, \\
\hat{f}_{\mathcal{S}_2}(e) + \hat{f}_{\mathcal{S}_1}^{min}, \qquad e \in \Sigma_2 \setminus \Sigma_1,
\end{array}\right.
\end{equation}
\\
In the special case of stream summaries holding exactly $k = 2$ counters, it holds that for $i = 1, 2$, $\left\vert\Sigma_i\right\vert \leq 2$, and $\left\vert\Sigma_C\right\vert \leq 4$. Now, suppose that $\left\vert\mathcal{S}_i\right\vert = \left\vert\mathcal{N}_i\right\vert$, (this is true when $\mathcal{S}_i$ is produced by the sequential Space Saving) and let $\delta = \hat{f}_{\mathcal{S}_1}^{min} + \hat{f}_{\mathcal{S}_2}^{min}$ and $x = \left\vert{\Sigma_C}\right\vert - 2$. Furthermore, suppose that the entries in $\mathcal{S}_C$ are sorted in ascending order with regard to the counters’ frequencies. \\
As proved in \cite{Cafaro-Pulimeno-Tempesta}, it holds that:

\begin{equation}
\label{eq10}
\left\vert{\mathcal{S}_C}\right\vert = \left\vert{\mathcal{S}_1}\right\vert + \left\vert{\mathcal{S}_2}\right\vert + x \delta = \left\vert{\mathcal{N}_1}\right\vert + \left\vert{\mathcal{N}_2}\right\vert + x \delta
\end{equation}

\begin{equation}
\label{eq22}
\left\vert{\mathcal{S}_M}\right\vert = \left\vert{\mathcal{N}_1}\right\vert + \left\vert{\mathcal{N}_2}\right\vert + x \delta - \sum_{i=1}^x \hat{f}_{\mathcal{S}_C}(e_i),
\end{equation}
\\
where the sum is extended over the first $x$ entries.

We have to show that the difference $x \delta - \sum_{i=1}^x \hat{f}_{\mathcal{S}_C}(e_i)$ is always equal to zero when $k = 2$, so that $\left\vert{\mathcal{S}_M}\right\vert = \left\vert{\mathcal{N}}\right\vert$.

When $x \leq 0$, $x \delta = 0$. In that case, $\mathcal{S}_M = \mathcal{S}_C$ and $\left\vert{\mathcal{S}_C}\right\vert = \left\vert{\mathcal{N}}\right\vert$.

When $x > 0$, the first $x$ counters of $\mathcal{S}_C$ have values equal to $\delta$. To see this, consider the two cases $x = 1$ and $x = 2$.

When $x = 2$, that is, the two summaries to be merged contain different items and $\left\vert{\Sigma_C}\right\vert = 4$, this is easily seen by simple computations: in fact, $\delta$ is the minimum value a counter in $\mathcal{S}_C$ can assume, and there are at least two counters with this value in $\mathcal{S}_C$, obtained combining the two counters with minimum value in $\mathcal{S}_1$ and $\mathcal{S}_2$. As a consequence these counters are the first two, and $x \delta - \sum_{i=1}^x \hat{f}_{\mathcal{S}_C}(e_i) = 0$.

When $x = 1$, one of the following cases arises: 
\begin{enumerate}

\item one of the summaries (without loss of generality, let us suppose it is $\mathcal{S}_1$)  contains two counters, the other summary ($\mathcal{S}_2$) contains only one counter and no item is in common between the summaries. In this case, $\delta$ is equal to the minimum counter in $\mathcal{S}_1$ since $\hat{f}_{\mathcal{S}_2}^{min}=0$, but it is also the minimum counter in $\mathcal{S}_C$, hence it holds that $\delta - \hat{f}_{\mathcal{S}_C}(e_1) = 0$

\item both summaries contain two counters and they have exactly an item in common. In this case we further have to distinguish three cases:
\begin{enumerate}
\item the item in common has the minimum frequency in both the summaries. The combined frequency of this item will be equal to $\delta$ which is the sum of the minimum frequencies of two summaries. Its combined frequency is also the minimum in $\mathcal{S}_C$, hence it holds that $\delta - \hat{f}_{\mathcal{S}_C}(e_1) = 0$;
\item the item in common has the maximum frequency in both the summaries. Its combined frequency is also the maximum value in $\mathcal{S}_C$, and $\mathcal{S}_C$ contains two distinct items with combined frequency equal to $\delta$ which is also the minimum in $\mathcal{S}_C$, hence $\delta - \hat{f}_{\mathcal{S}_C}(e_1) = 0$;
\item the item in common appears with minimum frequency in one summary (without loss of generality, let us suppose in $\mathcal{S}_1$) and with maximum frequency in the other summary ($\mathcal{S}_2$). The combined frequency of the item which appears with minimum frequency in $\mathcal{S}_2$ is equal to $\delta$ which again is the minimum frequency of the counters in $\mathcal{S}_C$, hence $\delta - \hat{f}_{\mathcal{S}_C}(e_1) = 0$.
\end{enumerate} 
\end{enumerate}

Taking into account that in all of the cases when $x=1$ the $\mathcal{S}_C$ contains at least one item whose combined frequency is equal to $\delta$, it holds that $x \delta - \sum_{i=1}^x \hat{f}_{\mathcal{S}_C}(e_i) = 0$.

\end{proof}

We have shown that all of the properties of a Space Saving summary of two counters are preserved by the merge procedure introduced in \cite{Cafaro-Pulimeno-Tempesta}. It suffices to guarantee that all of the properties stated for an FDCMSS sketch continue to hold after the parallel merge procedure depicted in the algorithm presented. In particular, it holds the property \ref{ss1}, which guarantees that a merged FDCMSS sketch continues to be 1-norm equivalent to a Count--Min sketch.

\section{Experimental results}
\label{results}

In this section, we report experimental results on synthetic datasets. Here, we thoroughly test our algorithm using an exponential decay function. All of the experiments have been carried out on the Galileo cluster machine kindly provided by CINECA in Italy. This machine is a linux CentOS 7.0 NeXtScale cluster with 516 compute nodes; each node is equipped with 2 2.40 GHz octa-core Intel Xeon CPUs E5-2630 v3, 128 GB RAM and 2 16 GB Intel Xeon Phi 7120P accelerators (available on 384 nodes only). High-Performance networking among the nodes is provided by Intel QDR (40Gb/s) Infiniband. All of the codes were compiled using the Intel C++ compiler v17.0.0.

Let $f$ be the true frequency of an item and $\hat{f}$ the corresponding frequency reported by an algorithm, then the Relative Error is defined as $\Delta f = \frac{{\left| {f - \hat{f}} \right|}}{f}$, and the Average Relative Error is derived by averaging the Relative Errors over all of the measured frequencies.

Precision, a metric defined as the total number of true heavy hitters reported over the total number of candidate items, quantifies the number of false positives reported by an algorithm in the output stream summary. Recall is the total number of true heavy hitters reported over the number of true heavy hitters given by an exact algorithm. In all of the results we obtained 100\% recall, even on a tiny sketch of size 4 x 800 (recall may be less than 100\%, but this happens only when the sketch size is really minimal). For this reason, to avoid wasting space, we do not show here recall plots. Rather, we present Precision, Absolute Error, Average Relative Error (ARE), Updates/ms  and runtime/performance plots since we are interested in understanding the error behavior and the algorithm's scalability when we use an increasing number of cores of execution. Table \ref{experiments} reports the experiments carried out. For each different metric under examination, we varied $n$, the stream size in billions of items, $\rho$, the skew of the zipfian distribution, $\phi$, the threshold and $w$, the number of sketch columns. All of the other parameters are fixed when varying one of the previous ones, and we show, on top of each plot, the fixed parameters' values.

Finally, we also present, for the metrics of interest, the results obtained by fixing the stream size and varying the number of cores utilized from 1 to 512. We conclude this section with a comparison between strong and weak scalability.

With the experiment 1 we aim at measuring the algorithm accuracy, the experiment 2 aims at measuring how the parallelization affects the algorithm's accuracy, finally experiment 3 is meant to measure the computational performance of the parallel algorithm measuring both strong and weak scalability.

\begin{table}[h]
 \caption{Design of experiments. The input stream size $n$ is expressed in billions, $\rho$ denotes the skewness of the input data distribution, $\phi$ is the support threshold, $w$ represents the number of columns in the sketch data structure, $p$ is the number of cores and $gs$ denotes the grain size, i.e. the number of elements (expressed in billions) in the input stream for each core. The \textit{nf} value stands for \textit{not fixed}}
      \label{experiments}
	\centering
    \begin{tabular}{|c|p{3.2cm}|c|c|c|c|c|c|c|}
    \hline
    Exp. & Aim & Varying & $n$ & $\rho$ & $\phi$ & $w$ & $p$ & $gs$ \\ \hline
    \multirow{4}{*}{1}  & \multirow{4}{\linewidth}{Algorithm accuracy} & $n=\{1, 2, 4, 8\}$ & nf & 1.1 & 0.01 & 1340 & 16 & 0.5 \\ \cline{3-9}
      &  & $\rho=\{1, 1.4, 1.8, 2.2\}$ & 8 & nf & 0.01 & 1340 & 16 & 0.5 \\ \cline{3-9}
      &  & $\phi=\{0.001, 0.004, 0.008, 0.016\}$ & 8 & 1.1 & nf & 1340 & 16 & 0.5 \\ \cline{3-9}
      &  & $w=\{800, 1600, 3200, 6400\}$ & 8 & 1.1 & 0.01 & nf & 16 & 0.5 \\ \hline \hline
     2 &  Parallel alg. accuracy & $p=\{1, 16, 128, 256, 512\}$ & 8 & 1.1 & 0.01 & 1340 & nf & nf  \\ \hline \hline
     \multirow{2}{*}{3}  & \multirow{2}{\linewidth}{Computational performance} & $p=\{1, 16, 128, 256, 512\}$ & 8 & 1.1 & 0.01 & 1340 & nf & nf \\ \cline{3-9}
      &  & $p=\{1, 16, 128, 256, 512\}$ & nf & 1.1 & 0.01 & 1340 & nf & 0.5\\ \hline
       \end{tabular}
    \end{table}

\subsection{Algorithm accuracy}
\begin{figure}[h]
  \centering
  \begin{tabular}{cc}  
  \subfloat[varying $n$]{
           \includegraphics[width=0.45\textwidth]{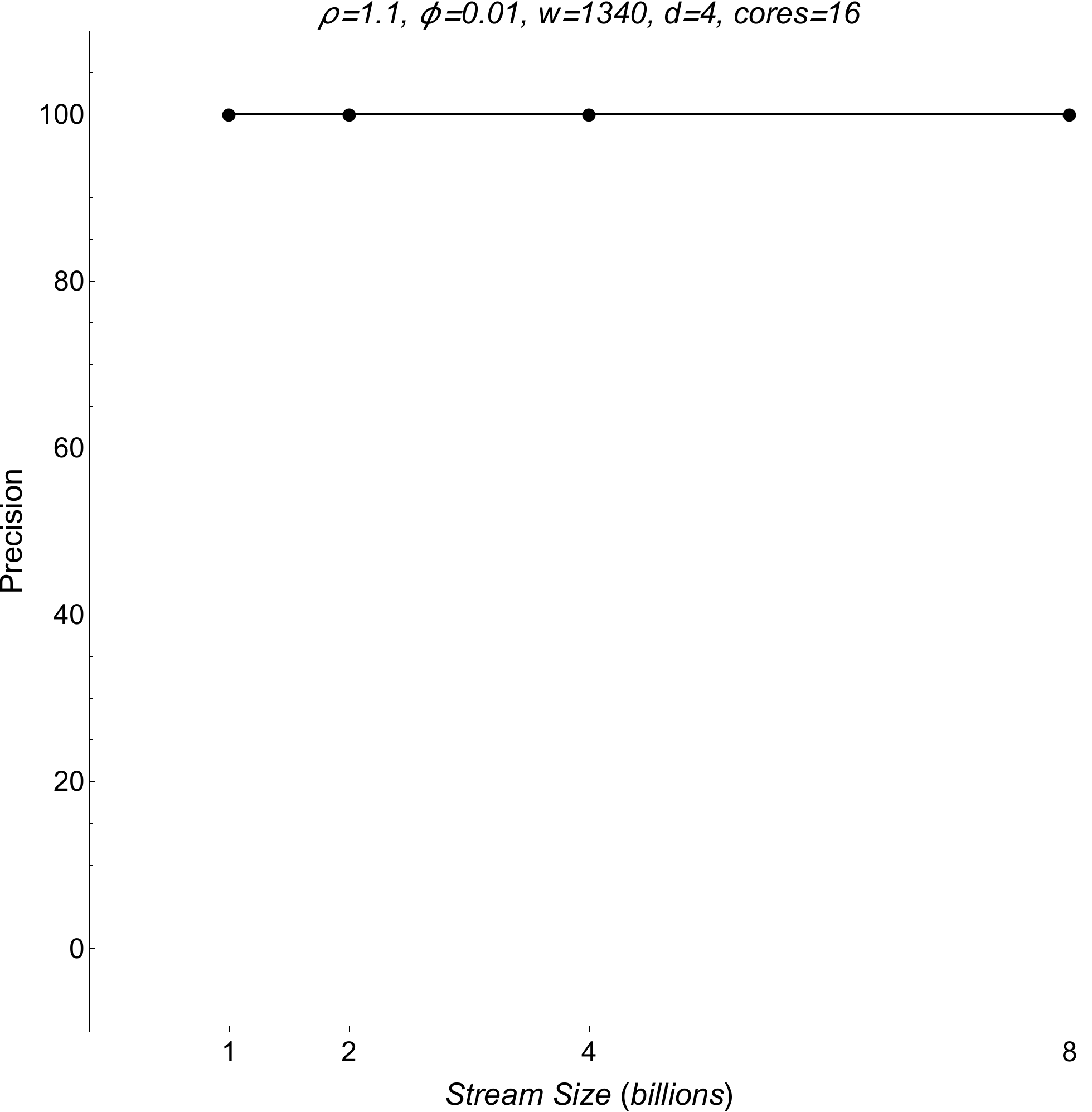}
           \label{ni-prec}
        } &
      \subfloat[varying $\rho$]{
           \includegraphics[width=0.45\textwidth]{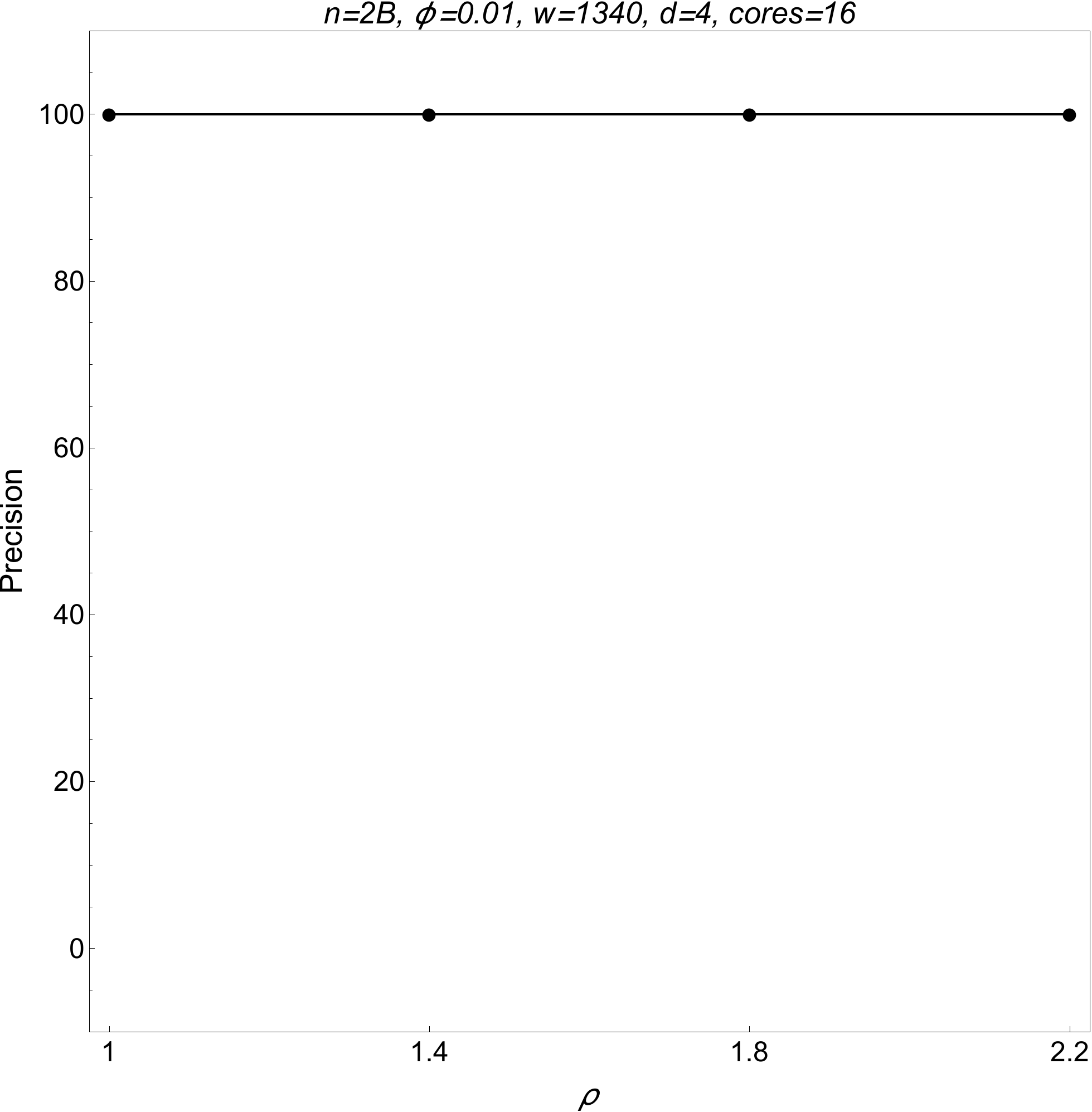}
           \label{sk-prec}
        } \\
      \subfloat[varying $\phi$ ]{
           \includegraphics[width=0.45\textwidth]{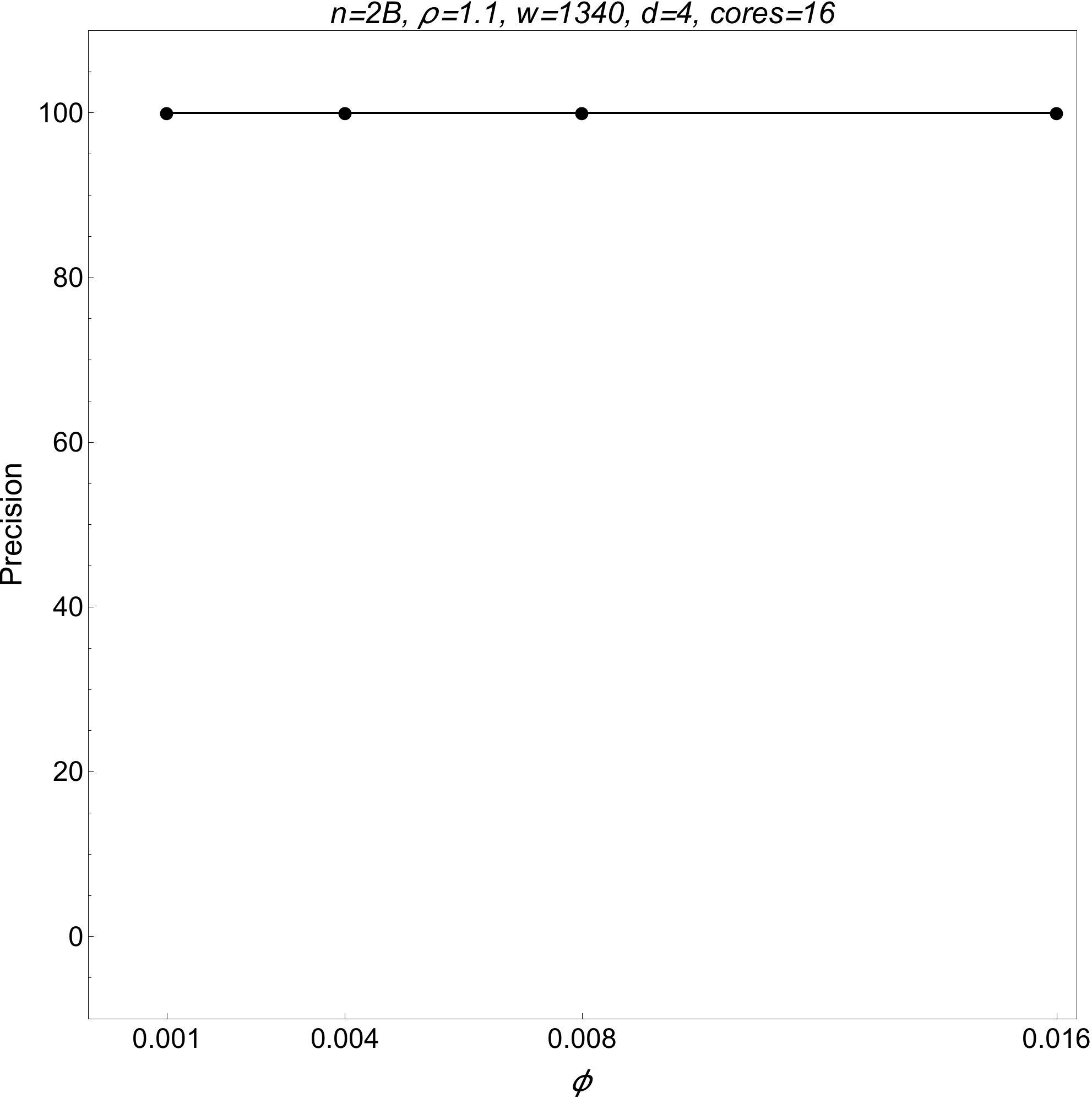}
           \label{thresh-prec}
        } & 
        \subfloat[varying $w$]{
           \includegraphics[width=0.45\textwidth]{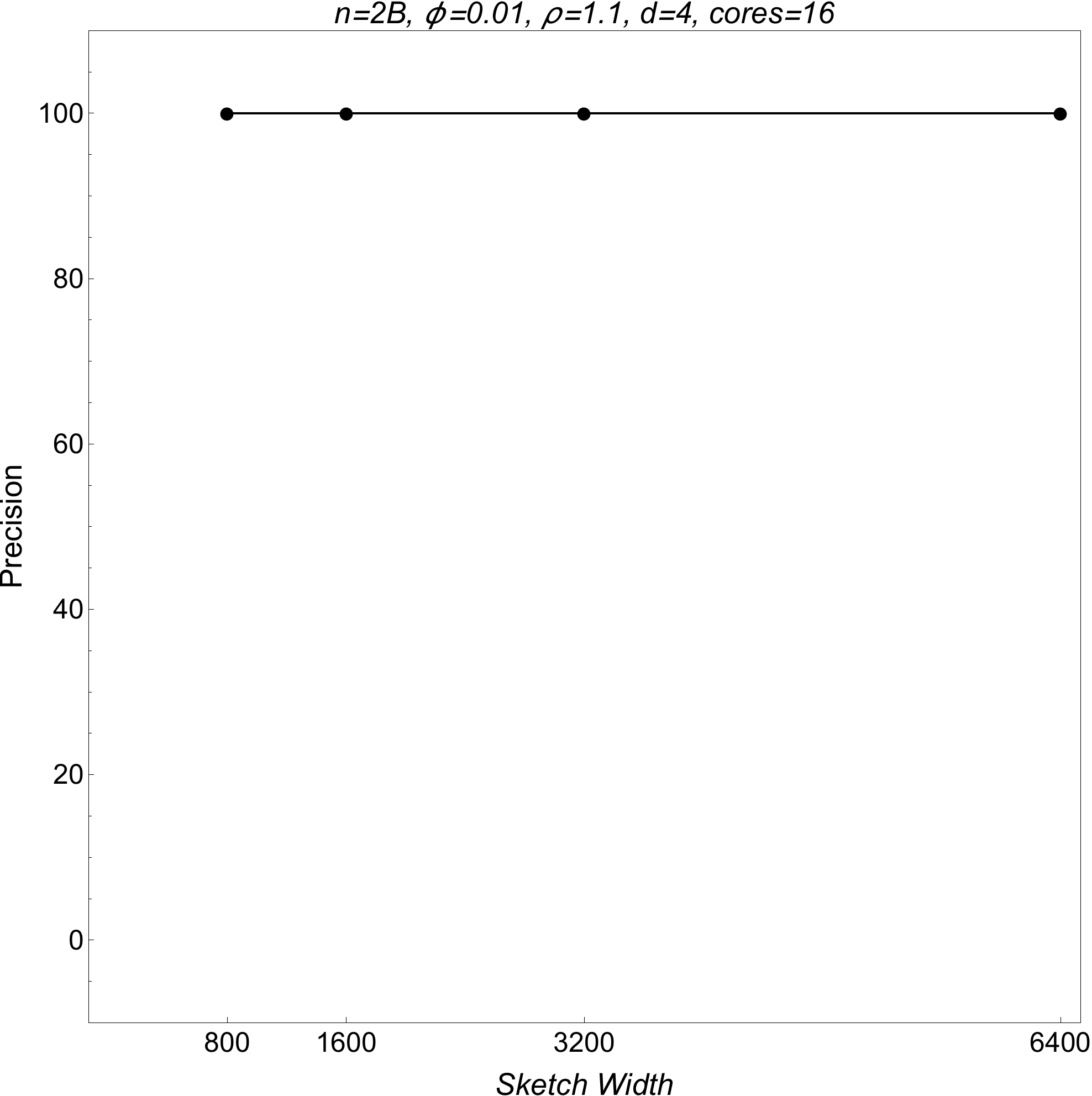}
           \label{width-prec}
        } 
\end{tabular}
 \caption{Precision} 
 \label{precision}
\end{figure}

\begin{figure}[h]
  \centering
  \begin{tabular}{cc}
  \subfloat[varying $n$]{
           \includegraphics[width=0.45\textwidth]{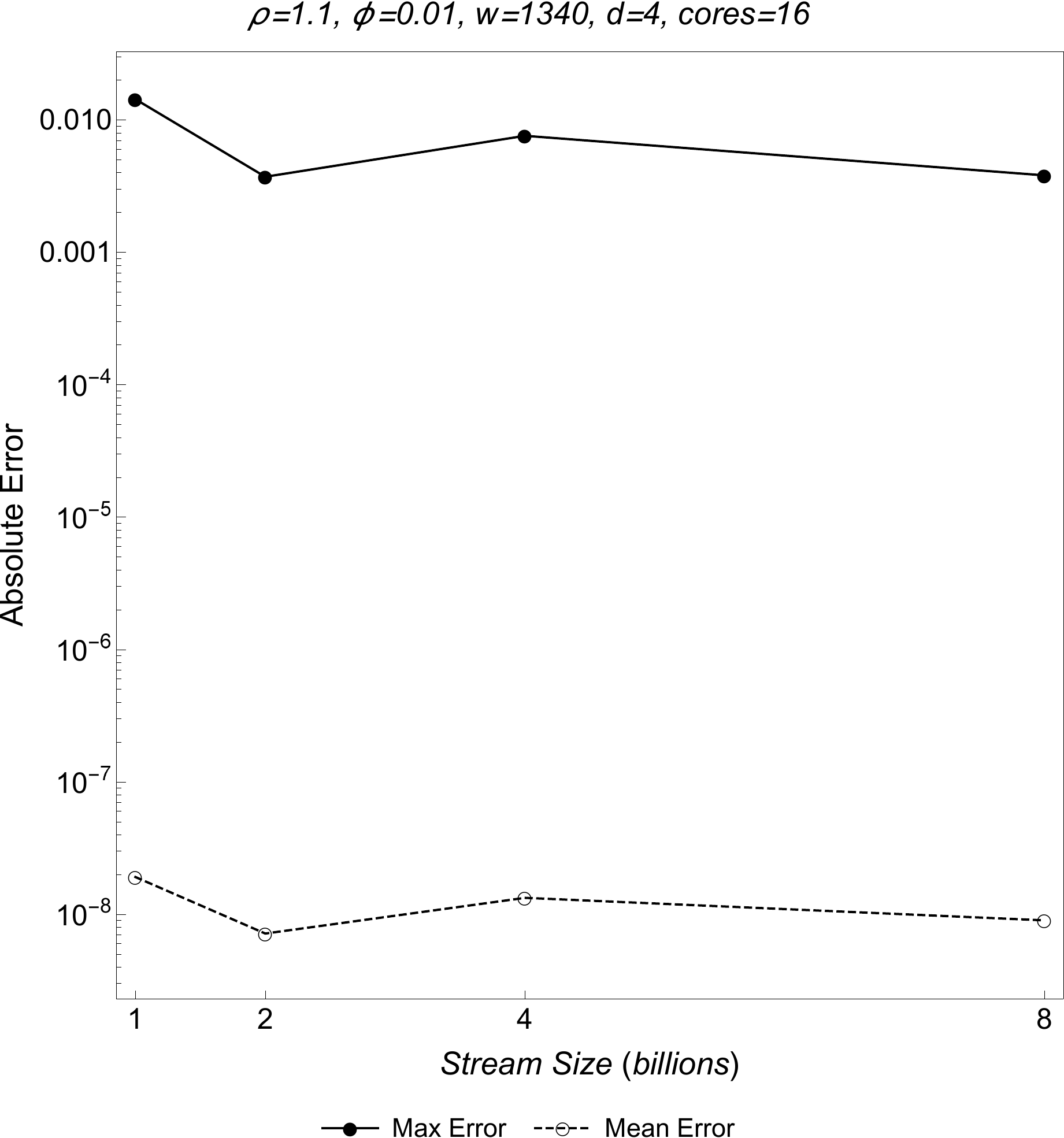}
           \label{ni-abserr}
        } &
      \subfloat[varying $\rho$]{
           \includegraphics[width=0.45\textwidth]{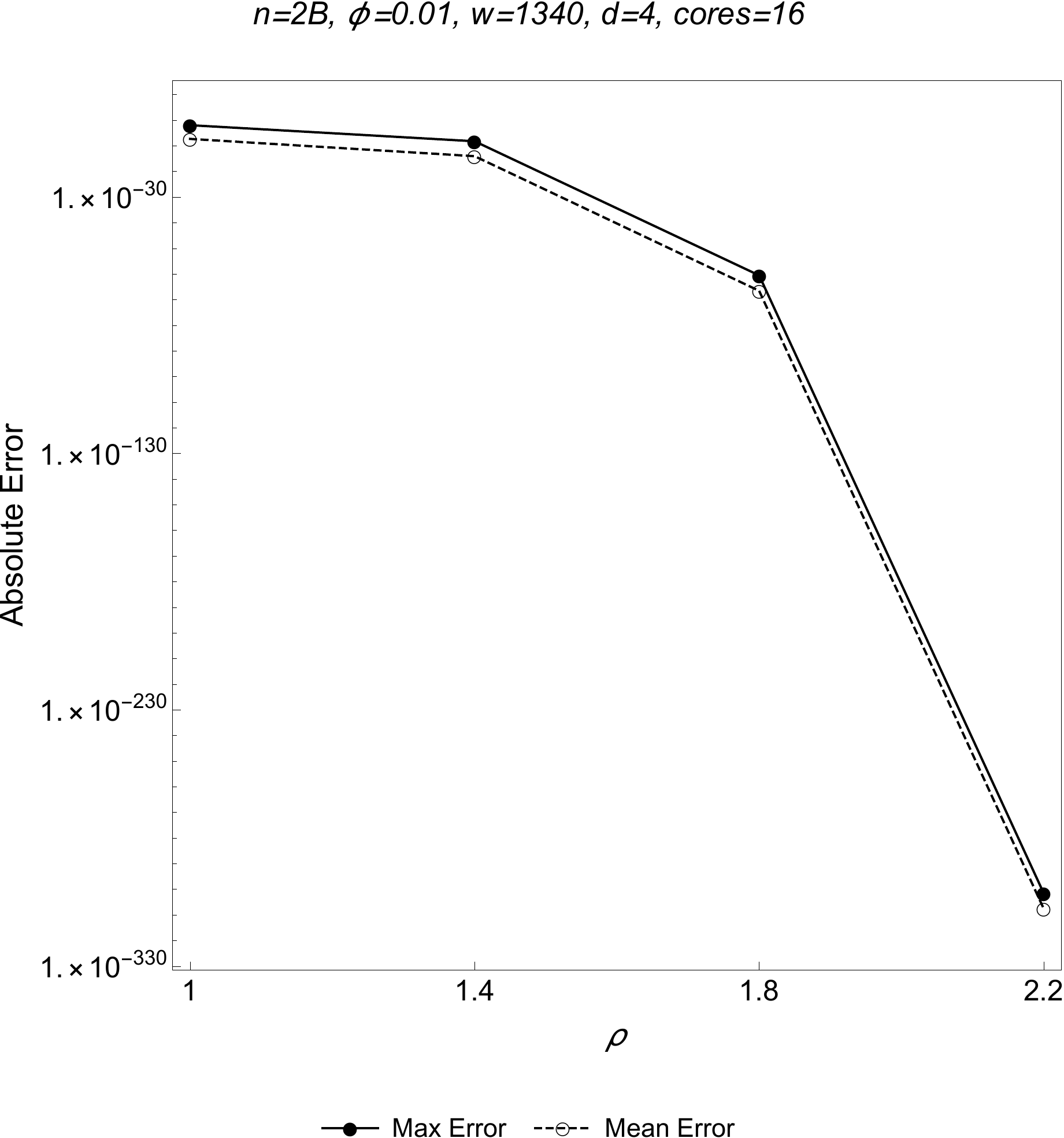}
           \label{sk-abserr}
        } \\
      \subfloat[varying $\phi$ ]{
           \includegraphics[width=0.45\textwidth]{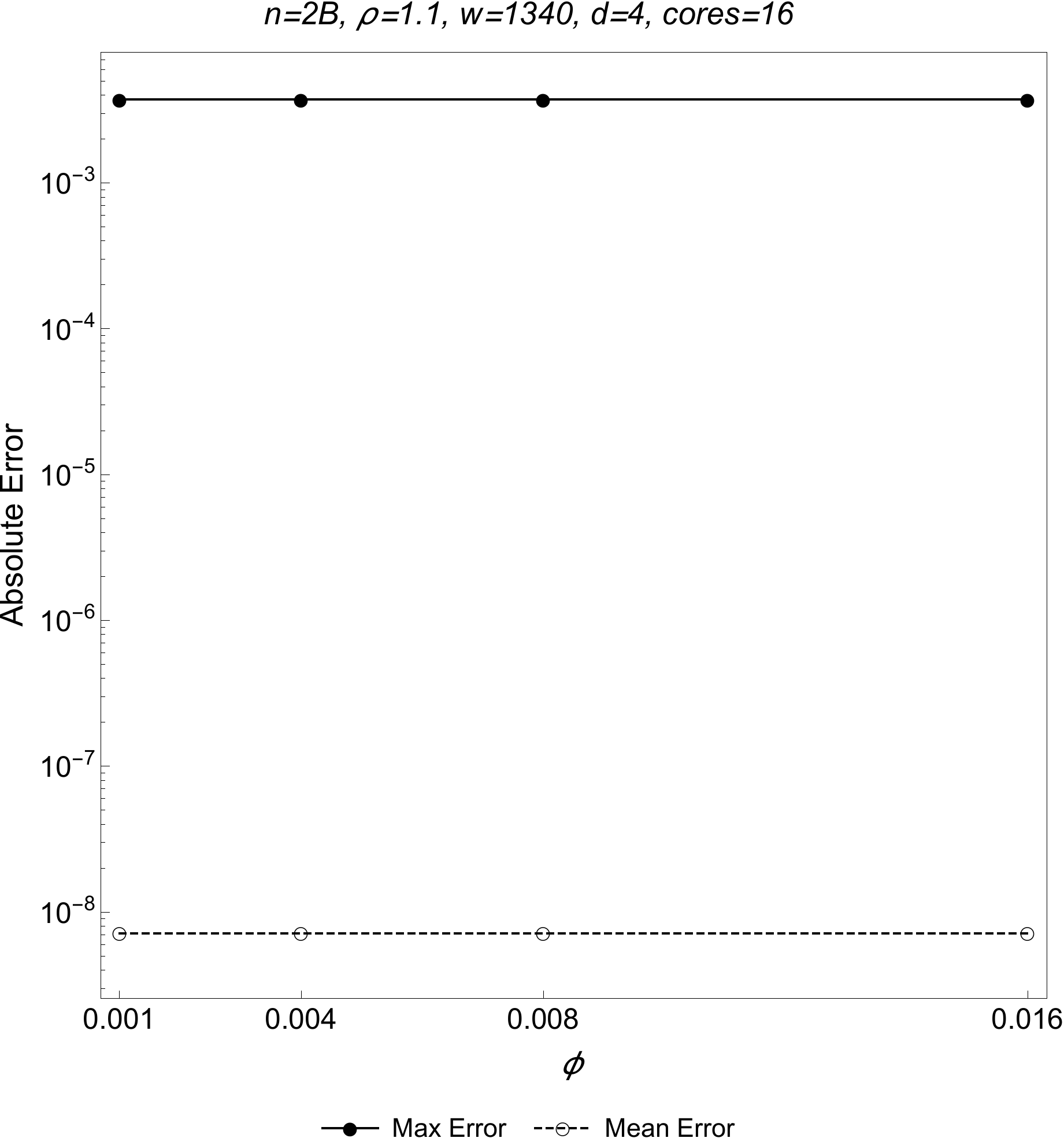}
           \label{thresh-abserr}
        } & 
        \subfloat[varying $w$]{
           \includegraphics[width=0.45\textwidth]{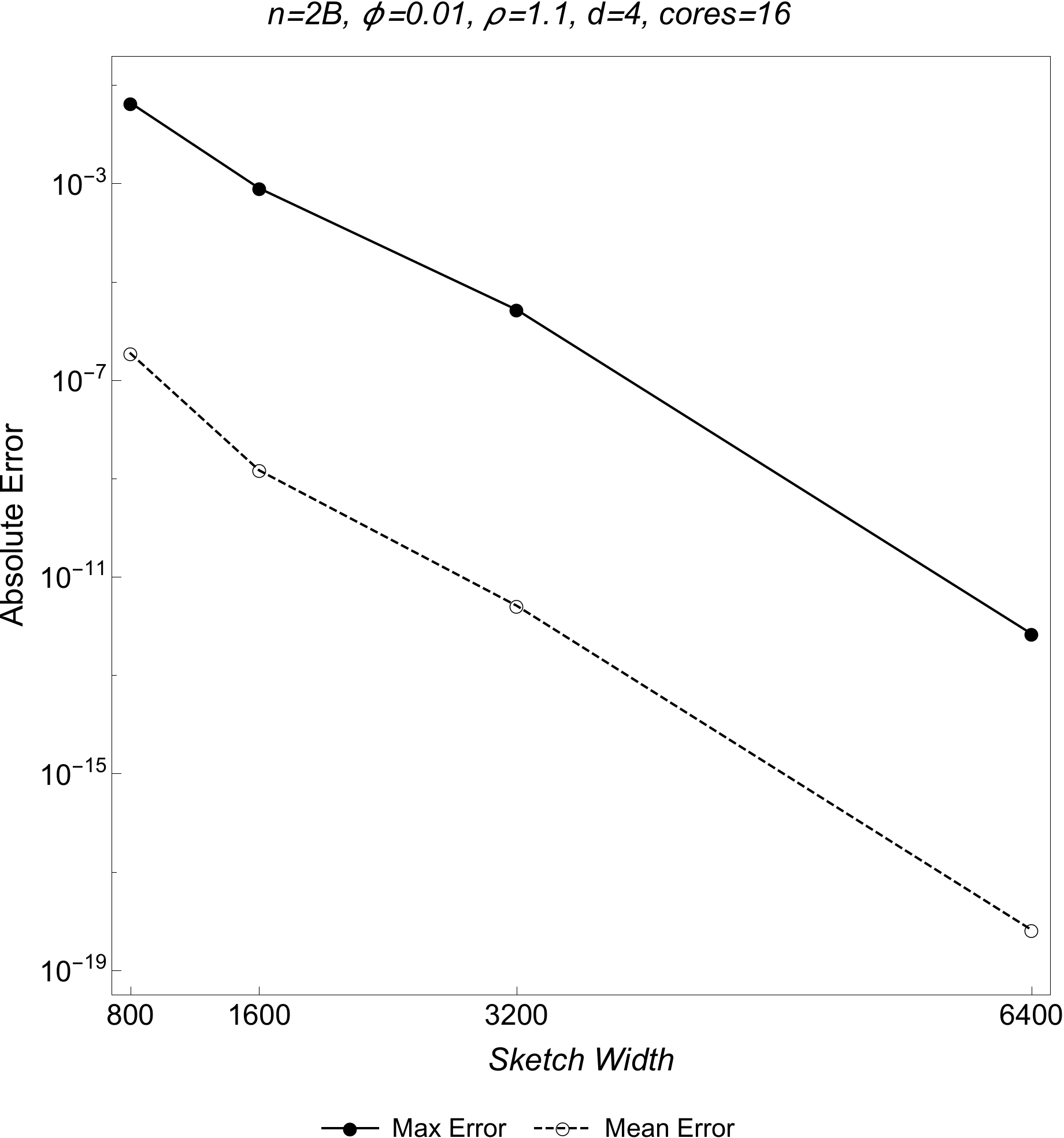}
           \label{width-abserr}
        } 
\end{tabular}
 \caption{Absolute Error} 
 \label{abserr}
\end{figure}

\begin{figure}[h]
  \centering
  \begin{tabular}{cc}
  \subfloat[varying $n$]{
           \includegraphics[width=0.45\textwidth]{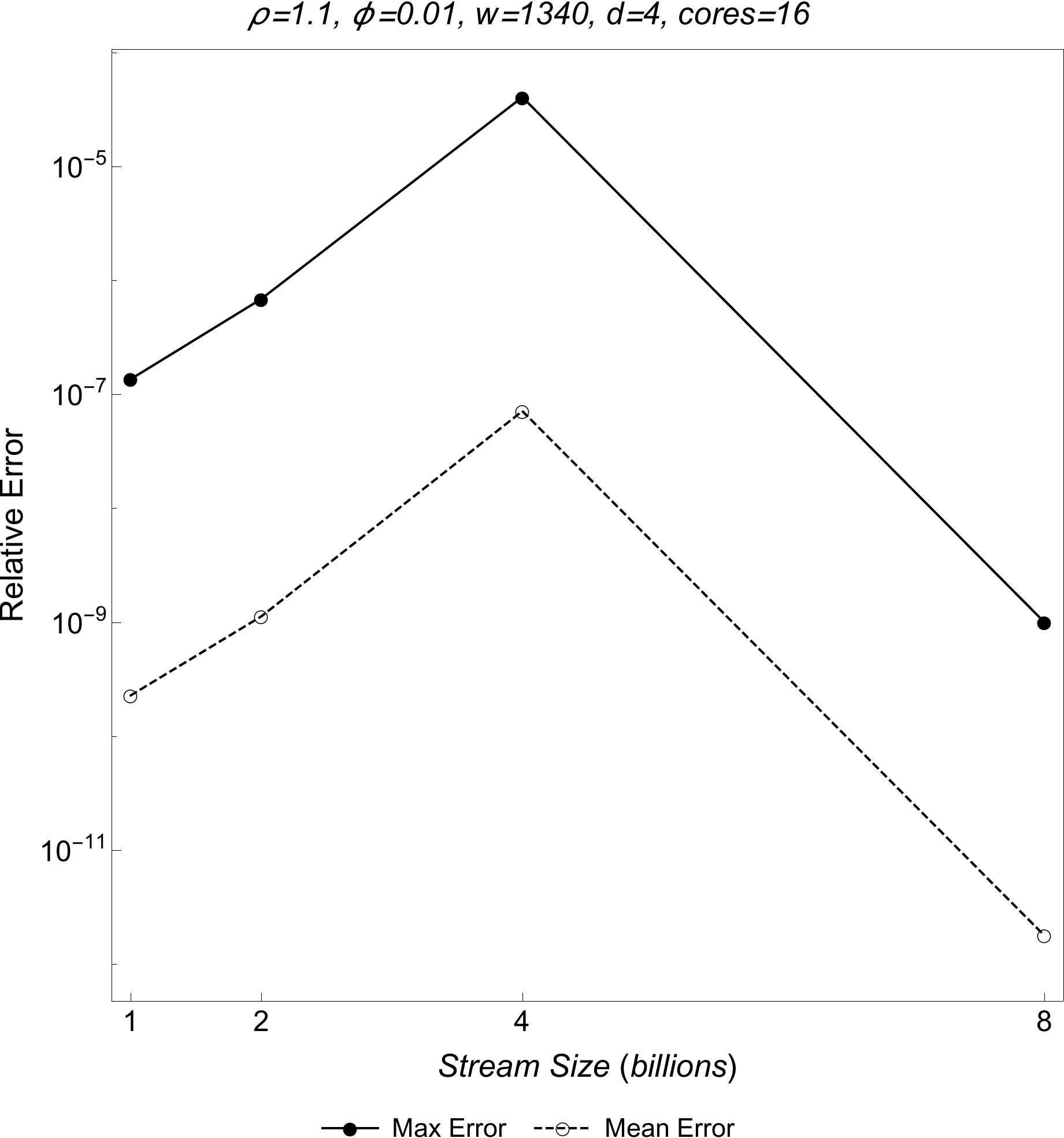}
           \label{ni-relerr}
        } &
      \subfloat[varying $\rho$]{
           \includegraphics[width=0.45\textwidth]{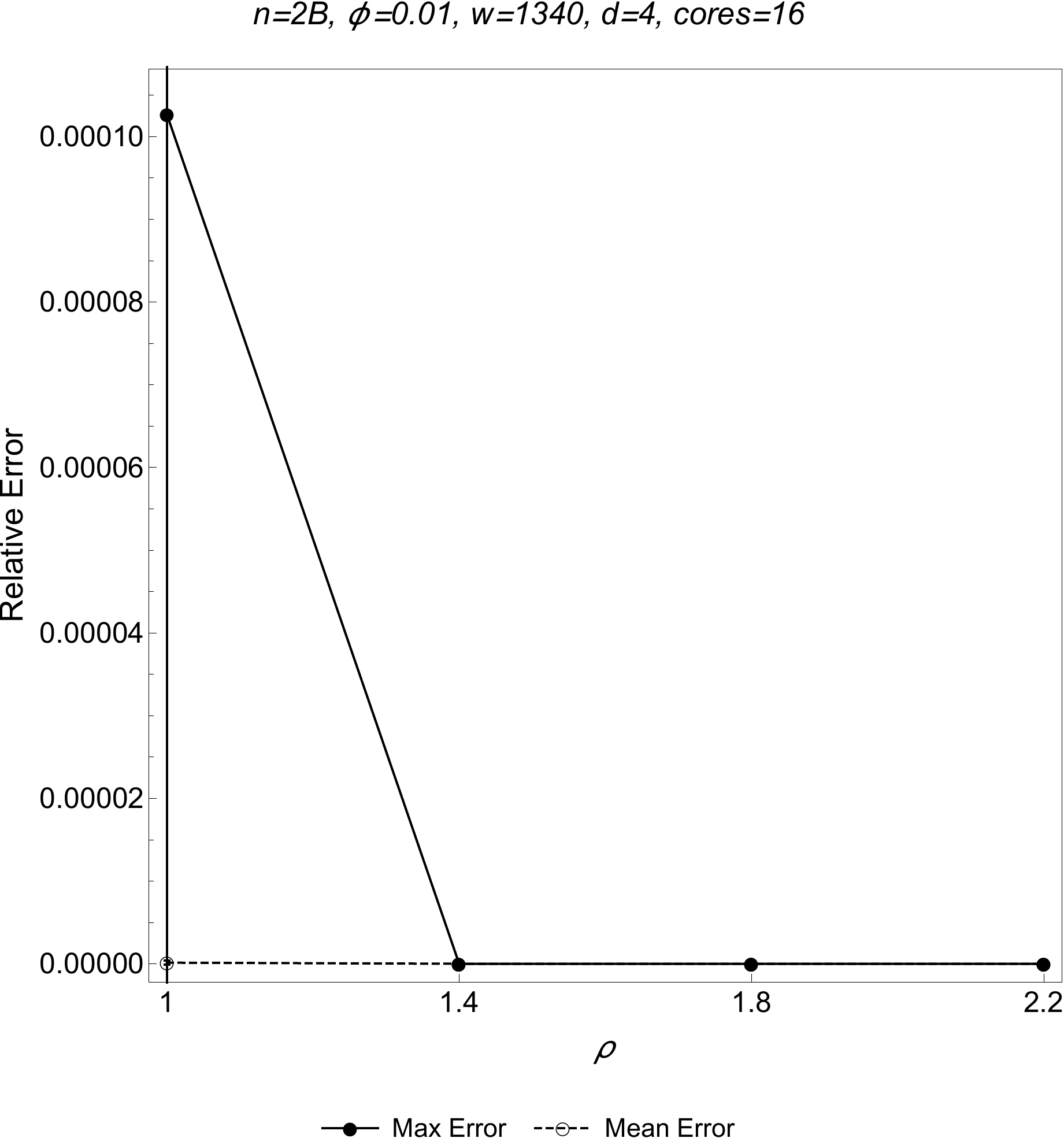}
           \label{sk-relerr}
        } \\
      \subfloat[varying $\phi$ ]{
           \includegraphics[width=0.45\textwidth]{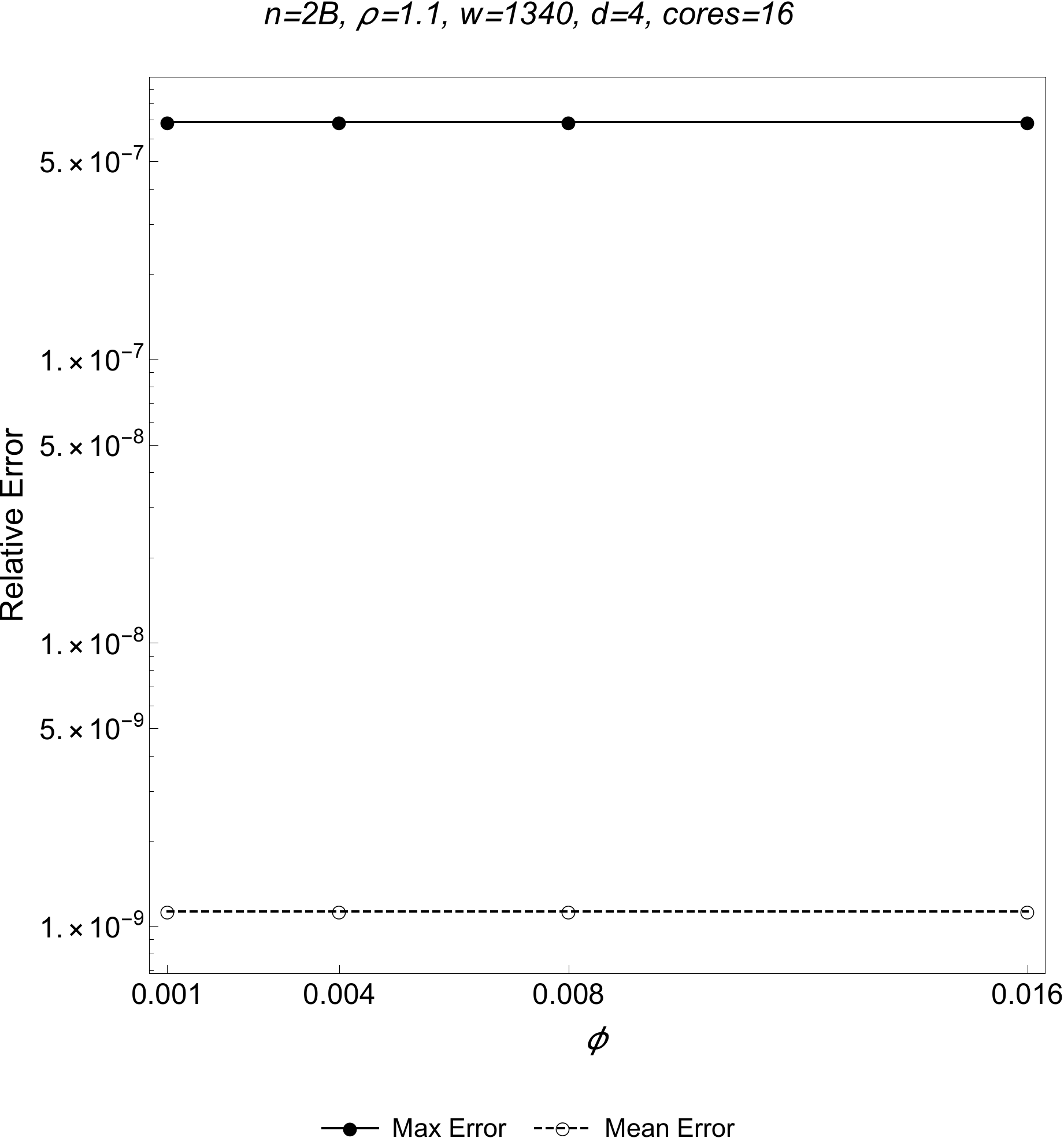}
           \label{thresh-relerr}
        } &
        \subfloat[varying $w$]{
           \includegraphics[width=0.45\textwidth]{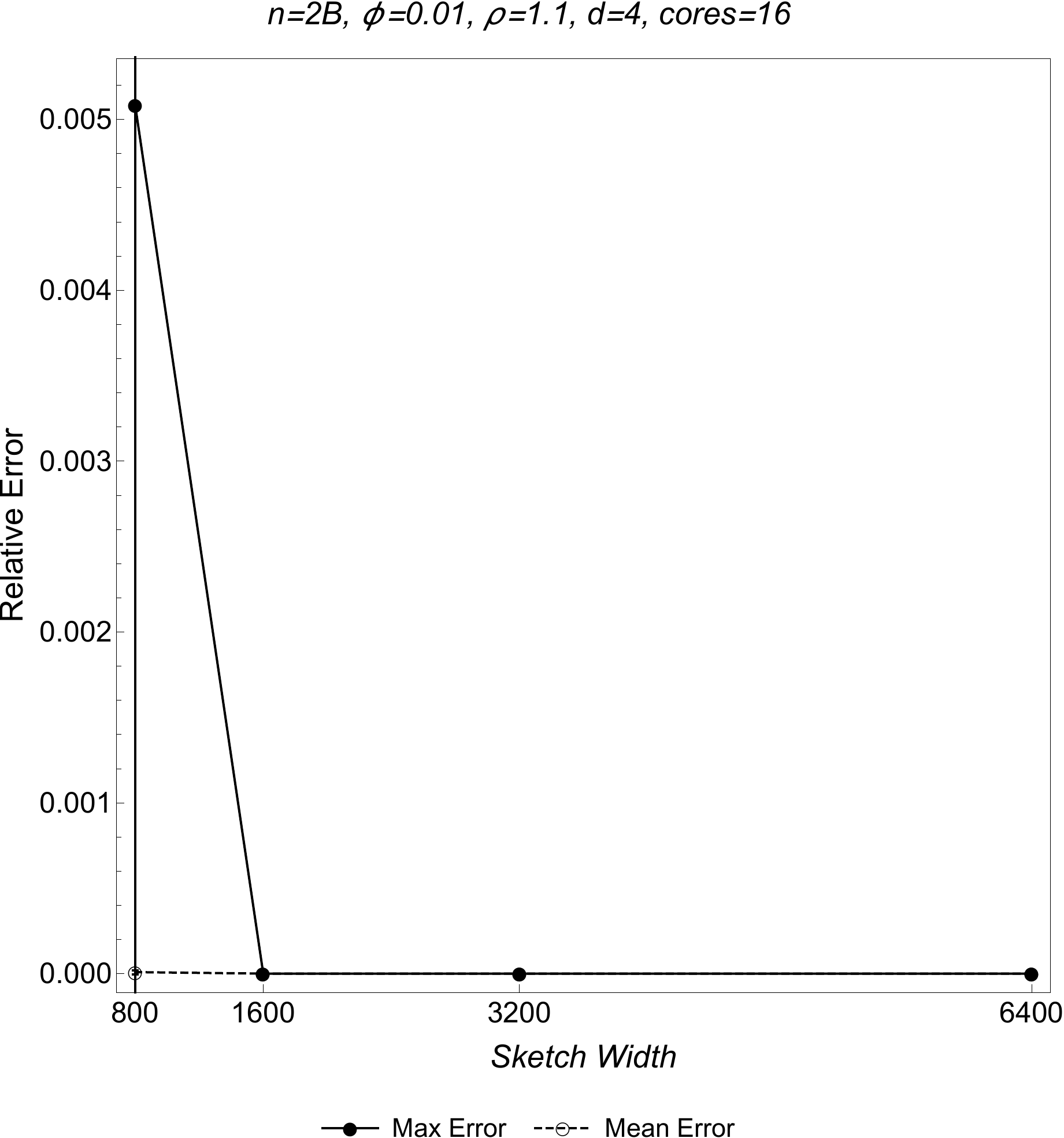}
           \label{width-relerr}
        } 
\end{tabular}
 \caption{Relative Error} 
 \label{relerr}
\end{figure}

As shown by Figure \ref{precision}, our parallel algorithm provides 100\% Precision in all of the experiments carried out. Both the Absolute and the Average Relative Error, depicted respectively in Figure \ref{abserr} and \ref{relerr}, have extremely low values, in particular with regard to their mean values. We observe that the Absolute Error is only slightly affected by the stream size $n$, and it's not affected at all by the threshold $\phi$. The behaviour observed when varying $\rho$ and $w$ is expected. We observe a decrement of the Absolute Error in both cases, since, when the skew is higher, the number of frequent items in the corresponding zipfian distribution is lower and, when $w$ is higher, increasing the sketch size provides better accuracy and, correspondingly, less error. Regarding the Average Relative Error, we observe the same qualitative behaviors in the experiments carried out.

\begin{figure}[h]
  \centering
  \begin{tabular}{cc}
  \subfloat[varying $n$]{
           \includegraphics[width=0.45\textwidth]{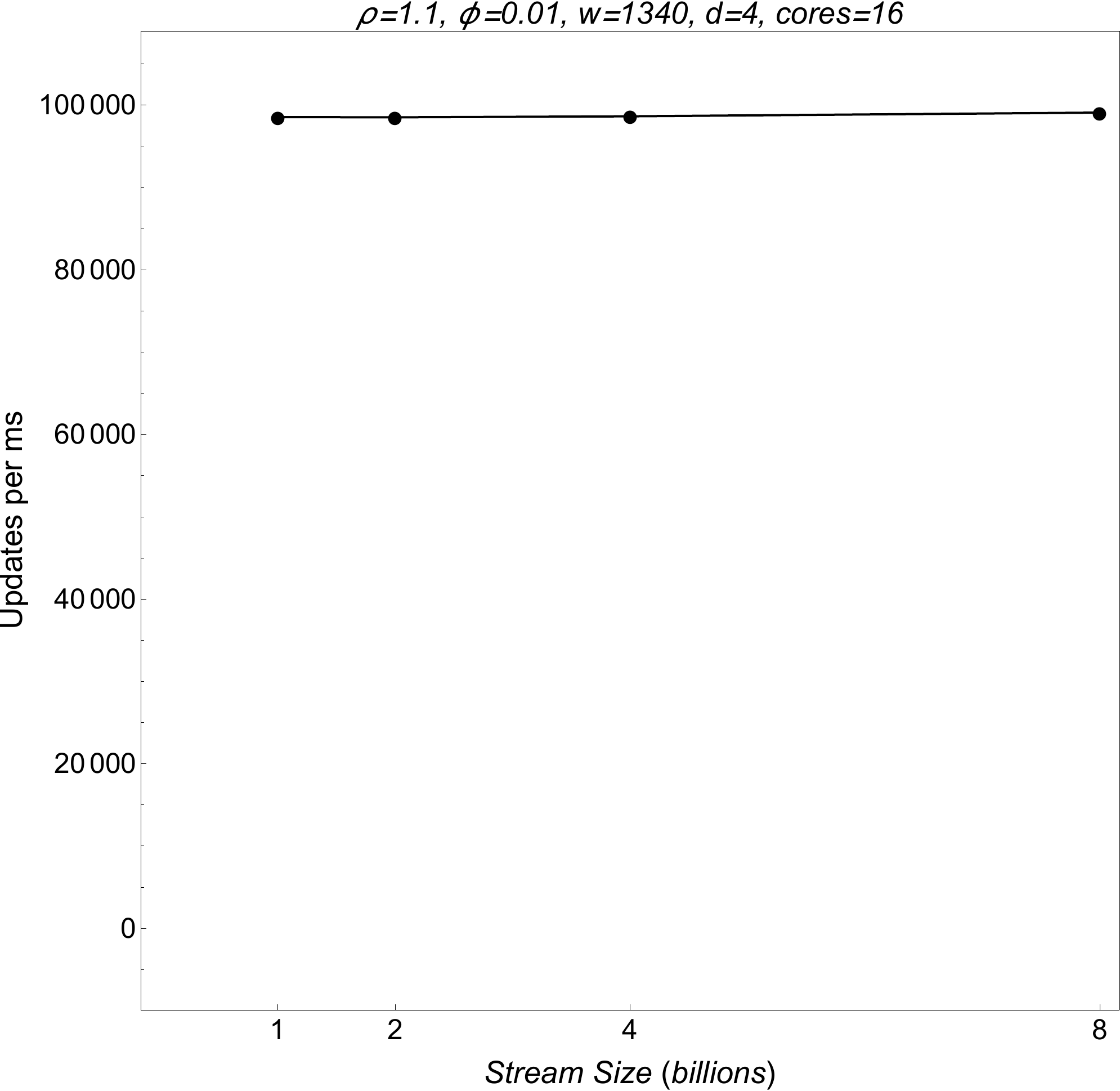}
           \label{ni-updates}
        } &
      \subfloat[varying $\rho$]{
           \includegraphics[width=0.45\textwidth]{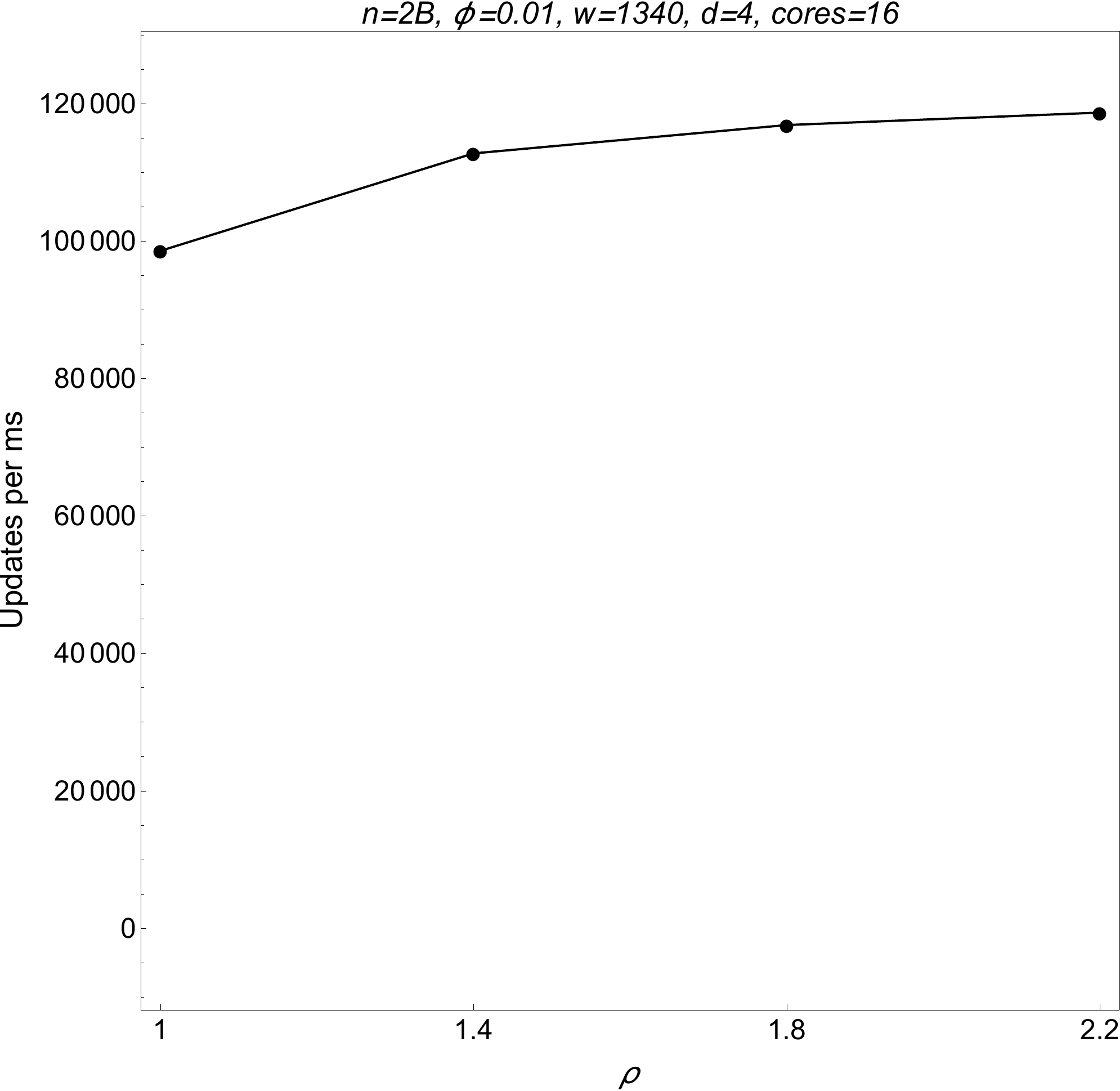}
           \label{sk-updates}
        } \\
      \subfloat[varying $\phi$ ]{
           \includegraphics[width=0.45\textwidth]{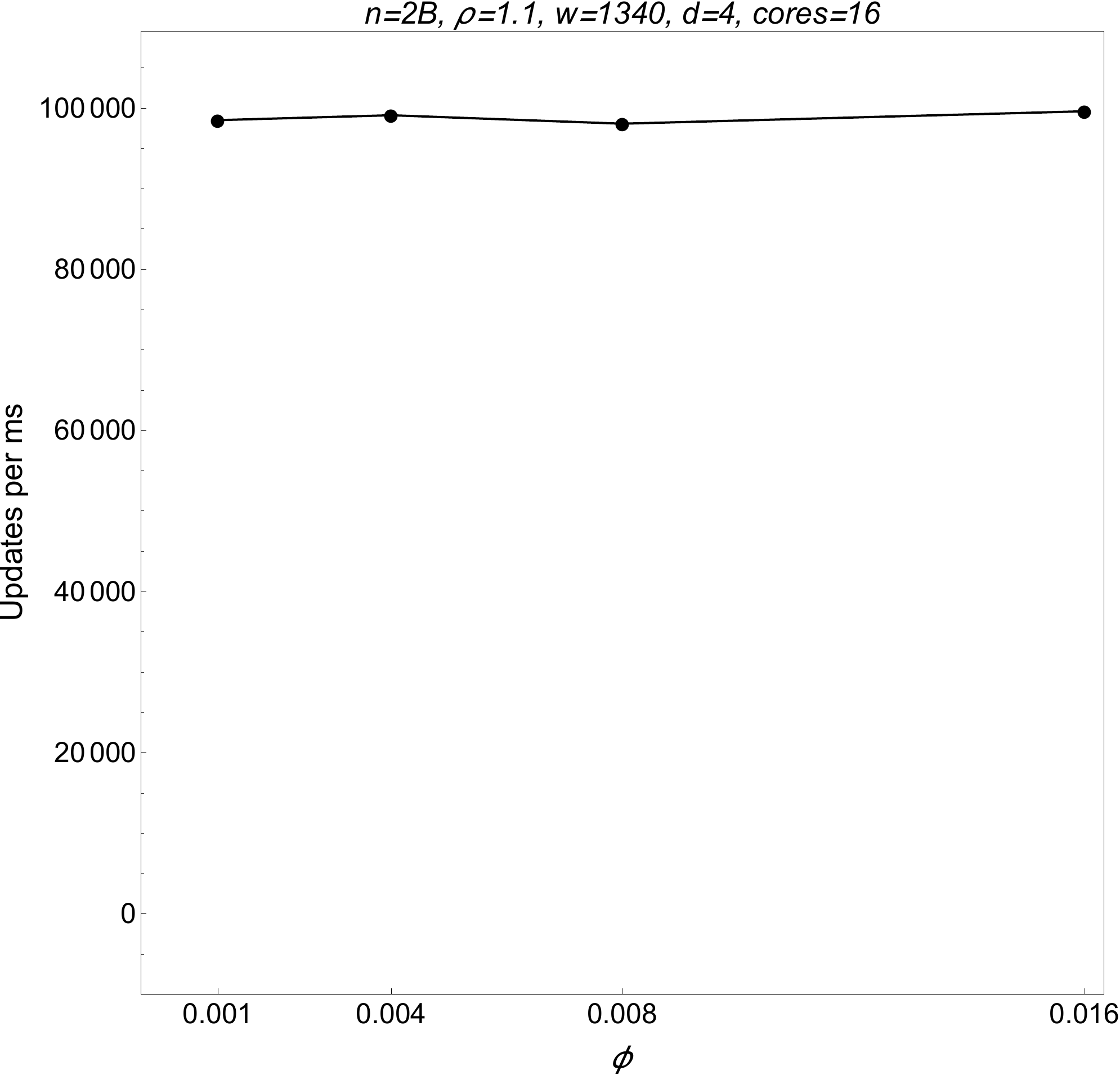}
           \label{thresh-updates}
        } &
        \subfloat[varying $w$]{
           \includegraphics[width=0.45\textwidth]{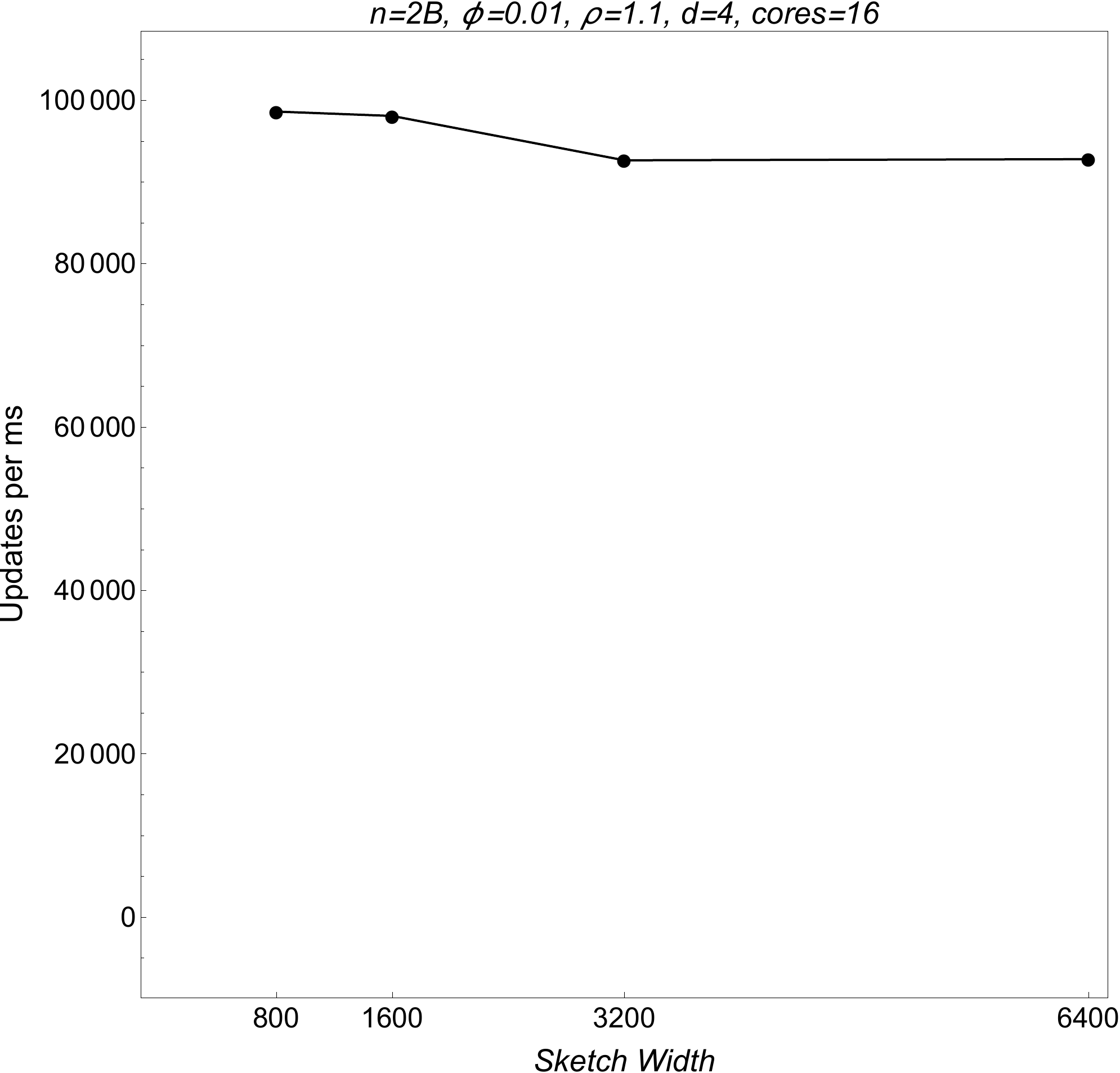}
           \label{width-updates}
        } 
\end{tabular}
 \caption{Updates} 
 \label{updates}
\end{figure}

Finally, the updates done per millisecond, shown in Figure \ref{updates}, appear to be stable around 100,000 when varying the stream size $n$ and the threshold $\phi$. There is a visible increment (from 100,000 to 120,000) when varying $\rho$ and a decrement  (from 100,000 to 90,000) when varying $w$. These behaviors are expected for the same reasons we gave when analyzing the error. Indeed, processing a stream in which the number of frequent items is lower is usually faster. On the other hand, increasing the sketch size provides better accuracy, but more time is required to update the sketch.  

\subsection{Impact of the parallelization on the accuray}
\begin{figure}[h]
  \centering
  \begin{tabular}{cc}
  \subfloat[Precision]{
           \includegraphics[width=0.45\textwidth]{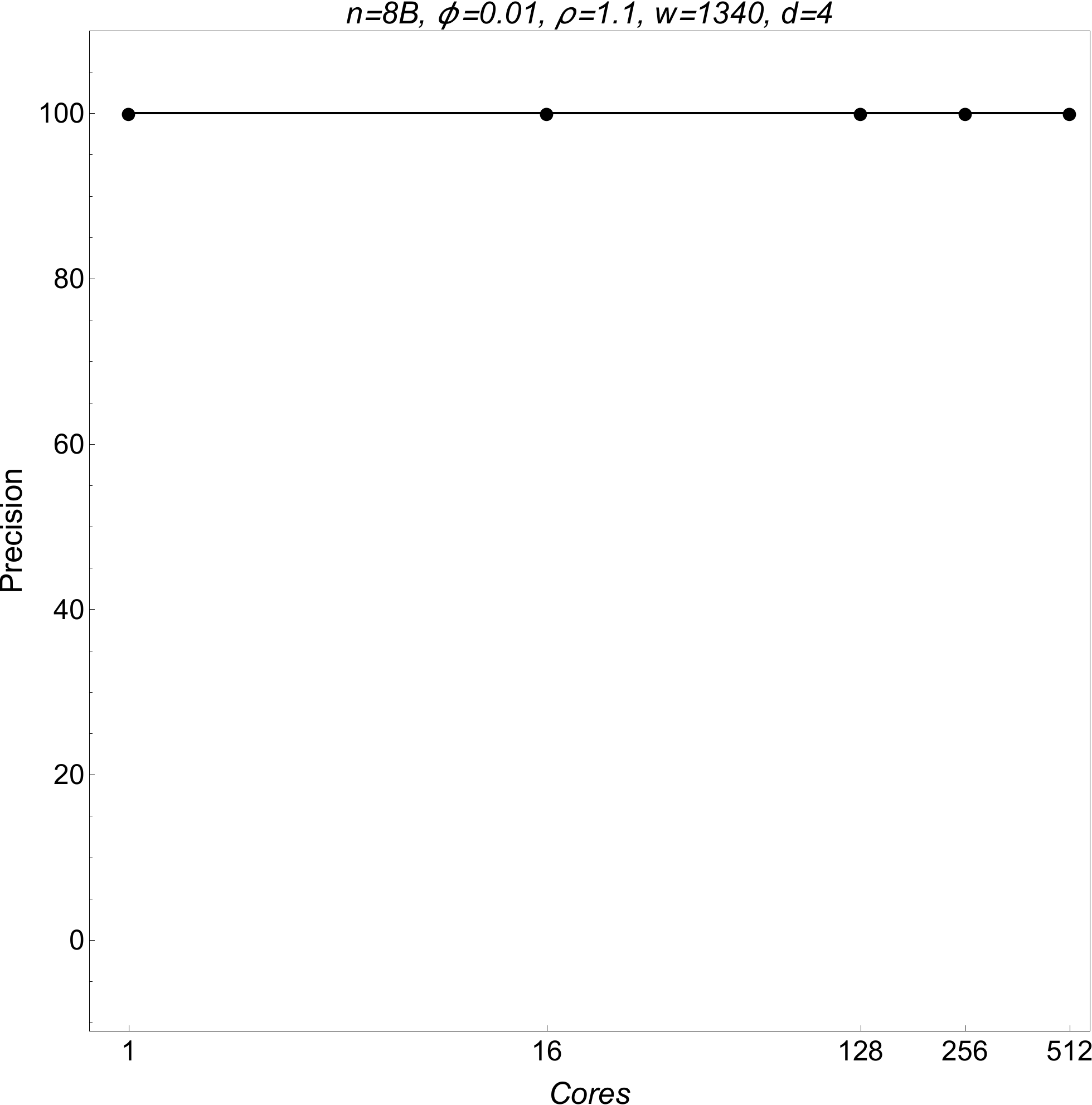}
           \label{prec-scalab}
        } &
        \subfloat[Updates/ms]{
           \includegraphics[width=0.45\textwidth]{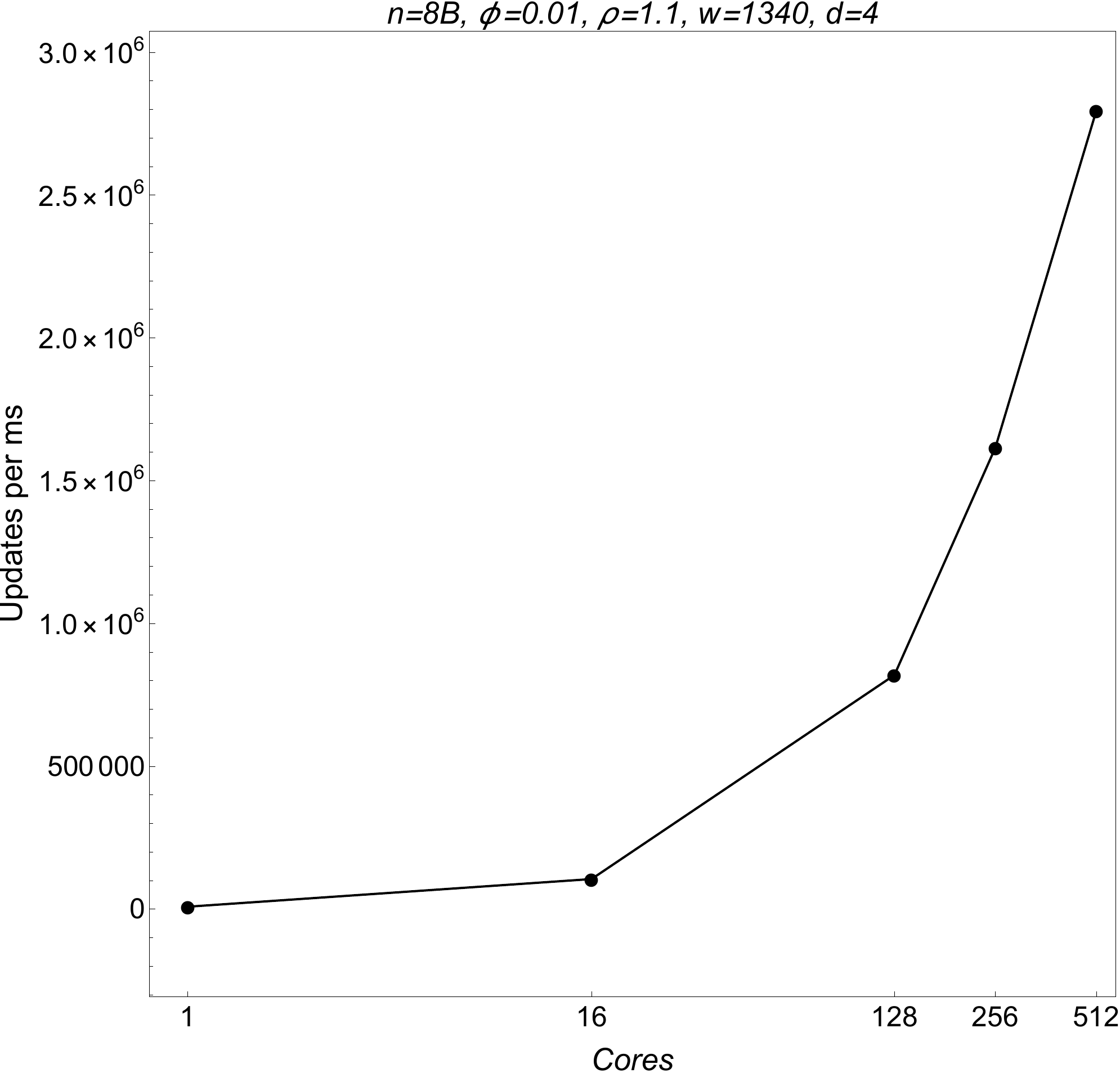}
           \label{updates-scalab}
        } \\
      \subfloat[Absolute Error]{
           \includegraphics[width=0.45\textwidth]{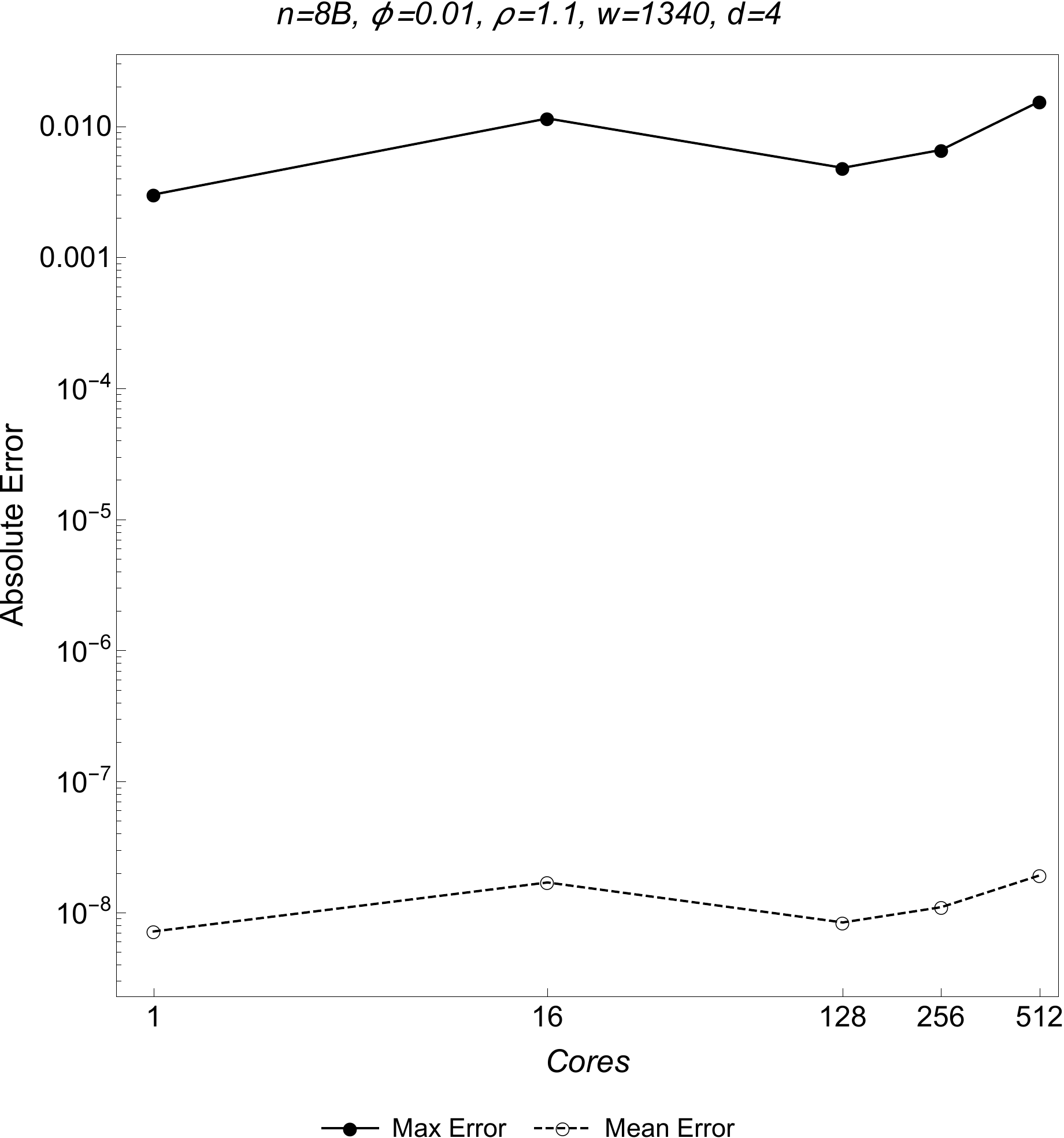}
           \label{abserr-scalab}
        } &
      \subfloat[Relative Error]{
           \includegraphics[width=0.45\textwidth]{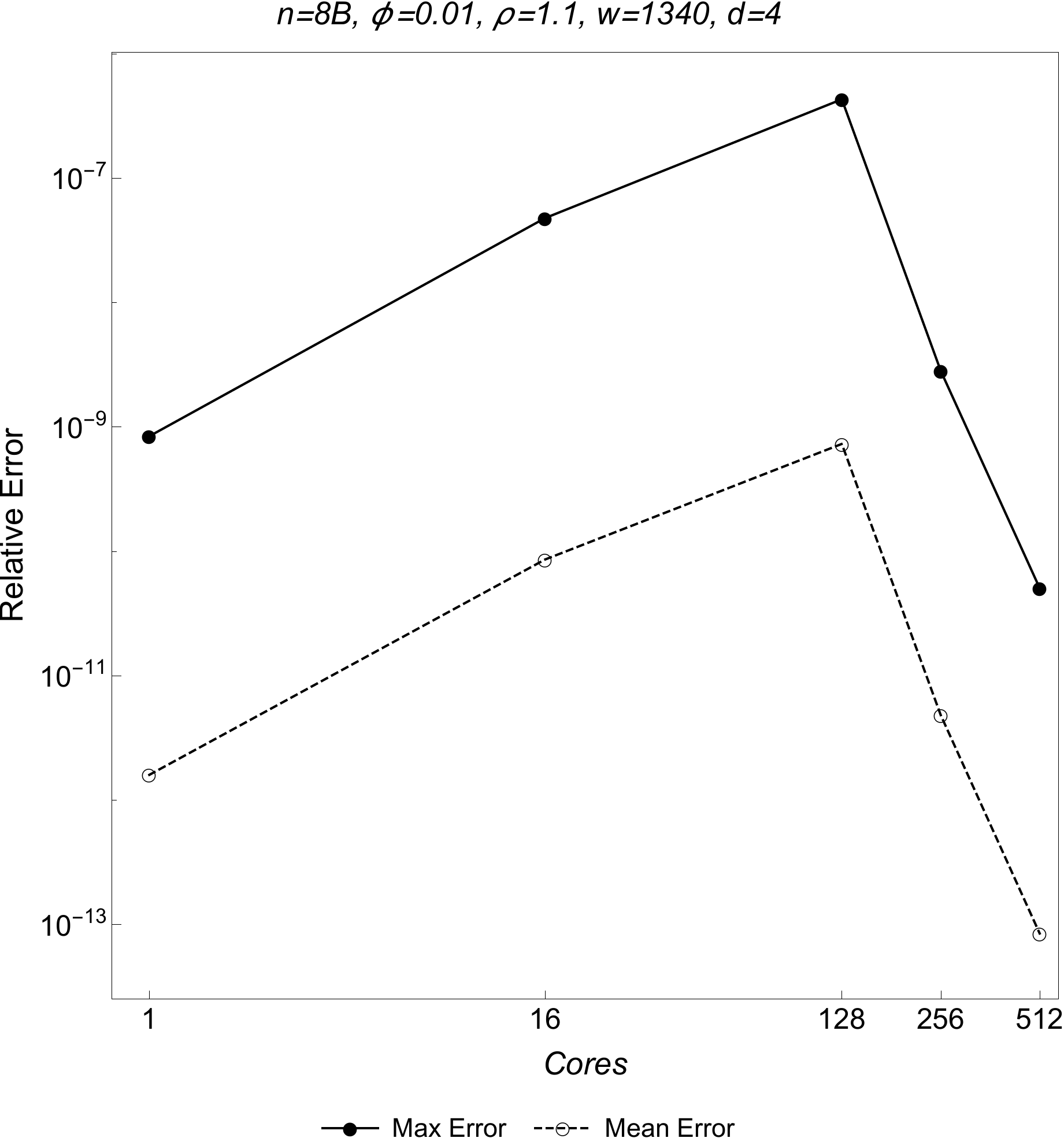}
           \label{relerr_scalab}
        }
\end{tabular}
 \caption{Scalability} 
 \label{scalability}
\end{figure}

We now discuss the experimental scalability shown by our parallel algorithm. Figure \ref{scalability} provides the results for the metrics under examination when testing strong scalability. That is, we fix the problem size (i.e., the stream size $n$) and increase the number of cores on which the algorithm is executed. As shown, Precision, Absolute and Average Relative Error are not affected at all by a strong scaling of the application, with Precision always equal to 100\% and extremely low error values. Finally, the observed increment of the Updates/ms when varying the number of cores utilized is expected, due to the frequency updates made in parallel. Ideally, the throughput of the algorithm, measured as updates/ms, should increase with the same rate of increase of the number of MPI processes. 

\subsection{Computational perfomance}
\begin{figure}[hbt]
  \centering
  \begin{tabular}{cc}
  \subfloat[Strong Scalability]{
           \includegraphics[width=0.45\textwidth]{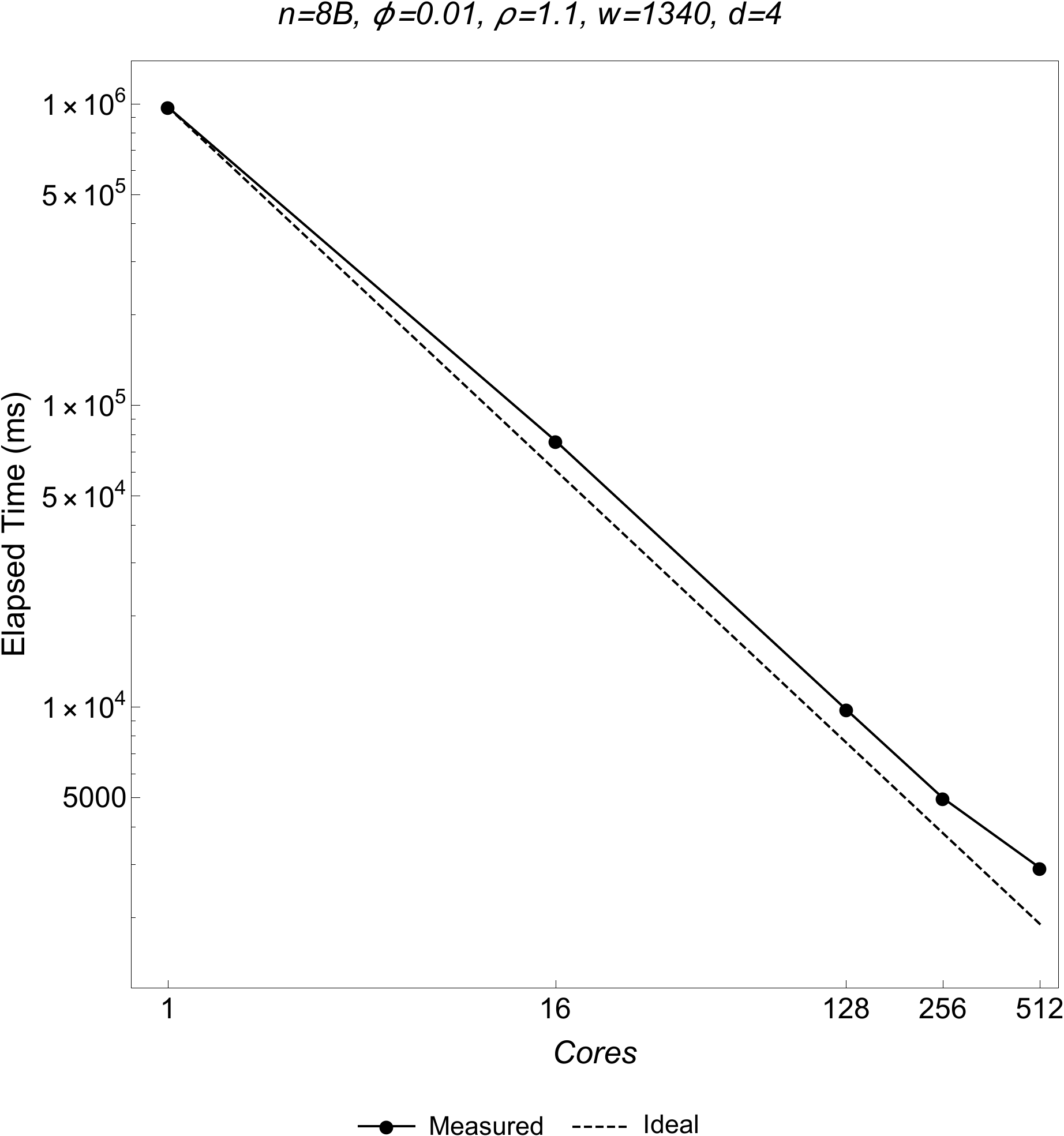}
           \label{prec-scalab}
        } &
      \subfloat[Weak Scalability]{
           \includegraphics[width=0.45\textwidth]{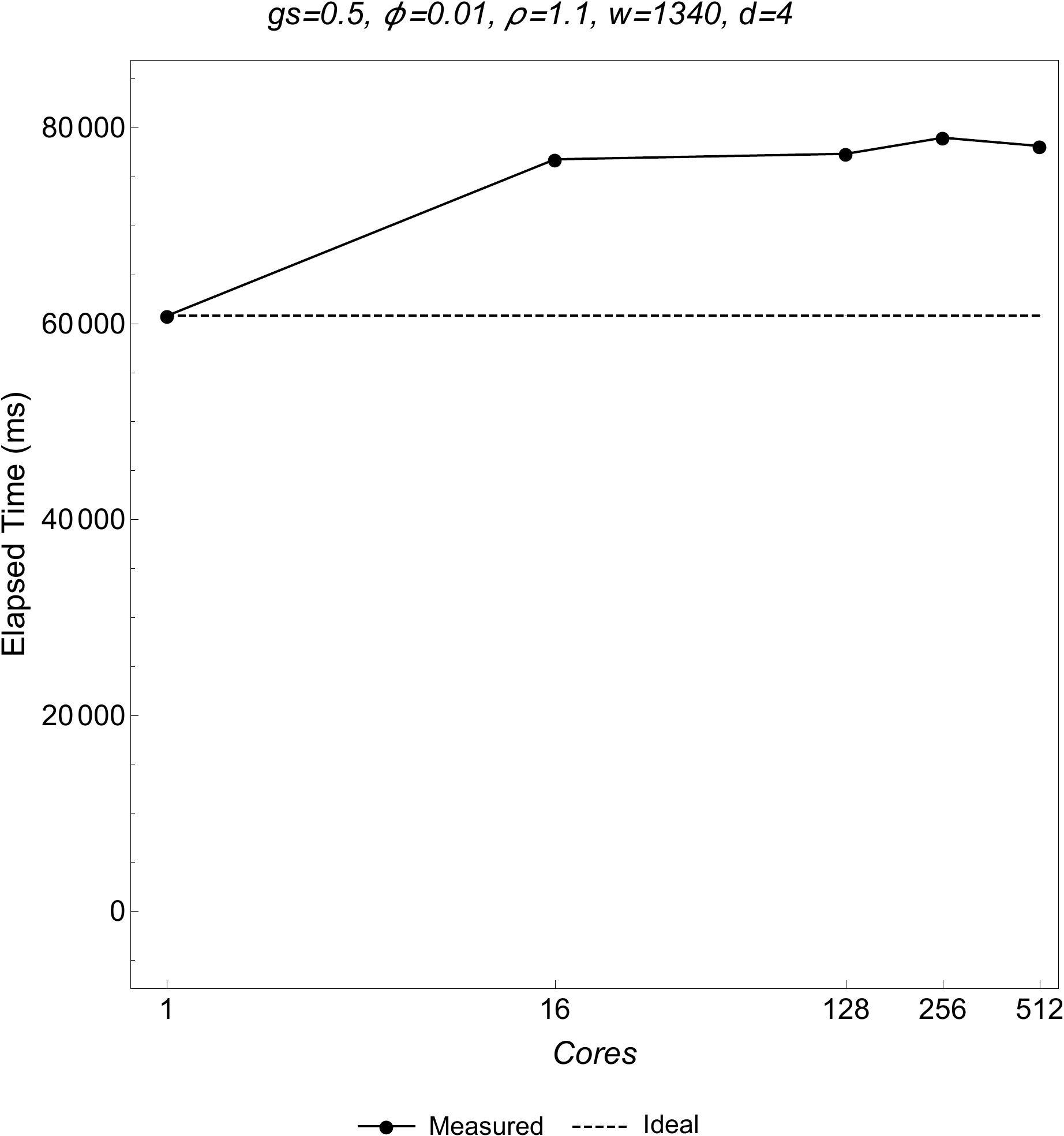}
           \label{abserr-scalab}
        } \\
\end{tabular}
 \caption{Parallel Scalability} 
 \label{strong-weak-scalability}
\end{figure}

Figure \ref{strong-weak-scalability} is related to the comparison we did to test weak scalability, which refers to the scalability of a parallel application obtained when the problem size is increased along with the number of cores, so that we measure how the running time changes with regard to the number of cores for a fixed problem size per core (whilst, for strong scalability, we measure how the running time changes with regard to the number of cores for a fixed total problem size). 

As shown, the plot for strong scaling is a log-log plot of the running time versus the number of cores. The dashed straight line with slope -1 indicates ideal scalability, whereas any upward curvature away from that line indicates limited scalability. The plot reports a good strong scalability even on 512 cores; this is due to the high number of items to be processed in the input stream which makes the computational time higher than the parallel overhead. 

Regarding weak scalability, the corresponding plot provides an indication of loss of performance when scaling from 1 to 16 cores while we have a very good scalability from 16 cores up to 512. This can be explained considering the parallel architecture used for testing. One computing node is made of two octa-core Xeon processors, so that the cores share the main memory banks and the third level cache memory. In the weak scalability experiment the problem size increases linearly with the number of processes, hence also the total memory increases linearly; since from 1 to 16 cores we use only one computing node the memory contention between parallel processes increases. When varying from 16 to 512 cores we use several different computing nodes ranging from 1 to 32; each node runs 16 processes which compete for memory accesses as already discussed, hence the further slight loss of performance is due to the communication overhead.  

\section{Conclusions}
\label{conclusions}
We have presented PFDCMSS, a novel message--passing based parallel algorithm for mining time--faded heavy hitters, which, to the best of our knowledge, is the first parallel algorithm solving the problem on message--passing parallel architectures. We have formally proved its correctness by showing that the underlying data structure, is non trivially mergeable. However, the parallel algorithm is fast and simple to implement, and we have shown, through extensive experimental results, that PFDCMSS retains the extreme accuracy and error bound provided by FDCMSS whilst providing very good parallel scalability.

%%%%%%%%%%%%%%%%%%%%%%%%%%%%%%%%%%%%%%%%%%%%%%%%%%%%%%%%%%%
% the following \clearpage command will prevent floats to appear in or after the references
\clearpage

\bibliographystyle{elsarticle-num}
\bibliography{bibliography}

% that's all folks
\end{document}